\documentclass[nohyperref,accepted]{article}

\usepackage{microtype}
\usepackage{graphicx}
\usepackage{subfigure}
\usepackage{booktabs} 

\usepackage{hyperref}

\usepackage{icml2022}

\usepackage{amsmath}
\usepackage{amssymb}
\usepackage{mathtools}
\usepackage{amsthm}

\usepackage[capitalize,noabbrev]{cleveref}

\theoremstyle{plain}

\theoremstyle{definition}

\theoremstyle{remark}

\usepackage[textsize=tiny]{todonotes}

\usepackage{amsfonts}
\usepackage{tikz}
\usetikzlibrary{quantikz}
\usepackage{array}
\usepackage{bm}
\usepackage{graphicx}
\usepackage{algorithm, algorithmic}

\icmltitlerunning{Quantum Discriminator for Binary Classification}

\begin{document}

\twocolumn[
\icmltitle{Quantum Discriminator for Binary Classification}



\icmlsetsymbol{equal}{*}

\begin{icmlauthorlist}
\icmlauthor{Prasanna Date}{ornl}
\icmlauthor{Wyatt Smith}{utk}
\end{icmlauthorlist}

\icmlaffiliation{ornl}{Oak Ridge National Laboratory, Oak Ridge, Tennessee, United States.}
\icmlaffiliation{utk}{University of Tennessee, Knoxville, Tennessee, United States}

\icmlcorrespondingauthor{Prasanna Date}{datepa@ornl.gov}

\icmlkeywords{Machine Learning, ICML}

\vskip 0.3in
]



\printAffiliationsAndNotice{}  

\begin{abstract}
Quantum computers have the unique ability to operate relatively quickly in high-dimensional spaces---this is sought to give them a competitive advantage over classical computers. 
In this work, we propose a novel quantum machine learning model called the Quantum Discriminator, which leverages the ability of quantum computers to operate in the high-dimensional spaces. 
The quantum discriminator is trained using a quantum-classical hybrid algorithm in $\mathcal{O}(N\log N)$ time, and inferencing is performed on a universal quantum computer in linear time. 
The quantum discriminator takes as input the binary features extracted from a given datum along with a prediction qubit, and outputs the predicted label. 
We analyze its performance on the Iris and Bars and Stripes data sets, and show that it can attain 99\% accuracy in simulation.

\end{abstract}

\section{Introduction}
\label{sec:intro}
Machine learning has become ubiquitous in almost every discipline under the sun \cite{qiu2016survey}. While high-quality training data will only continue to increase in availability in the coming decades, it is projected that classical approaches to machine learning will fail to keep pace with this increase owing to the end of Moore's Law \cite{aimone2019neural, preskill2018quantum}. Consequently, we must look towards alternative computing paradigms such as quantum and neuromorphic computing to address these scalability issues and develop more efficient machine learning methods \cite{biamonte2017quantum,date2021neuromorphic,date2019combinatorial}. 
    
Quantum computing has the potential to significantly speed up machine learning tasks \cite{biamonte2017quantum,ciliberto2018quantum}. 
Quantum computers use the quantum phenomena of superposition, tunneling and entanglement to perform computations \cite{nielsen2002quantum}.
As a result, they are able to operate in high-dimensional tensor-product spaces much faster than classical computers \cite{vazirani1998power,preskill1998quantum}.
For certain applications such as integer factoring and searching, quantum computers are known to outperform classical computers \cite{shor1994algorithms,grover1997quantum}.
We believe the ability of quantum computers to efficiently operate in these high-dimensional tensor product spaces can be leveraged to design efficient training and inferencing methods.

In this work, we focus on binary classification.
We operate within the traditional two-step workflow in machine learning, where the first step is to extract features from the data, and the second step is to perform classification using a discriminant function.
We further assume that binary features have been extracted from the data; many approaches for extracting binary features already exist in the literature \cite{ren2014face,chalkidis2017extracting}.
Therefore, our focus in this paper is to propose the Quantum Discriminator, which is a quantum discriminant model that performs binary classification on a set of binary features.
We also present the hybrid quantum-classical training algorithm used to train the quantum discriminator in $\mathcal{O}(N \log N)$ time.

As a proof of concept, we demonstrate that our model can be used to completely solve the $2$-bit binary classification problem. 
We also benchmark the quantum discriminator on the Iris and the Bars and Stripes data sets. 
Our results demonstrate that under a proper feature extraction and training regime, the quantum discriminator can attain a near-perfect ($99\%$) accuracy in simulation on the these data sets.  
Ideally, we would want to benchmark the quantum discriminator on benchmark data sets such as MNIST.
However, we are severely constrained by the small size and noisy nature of today's quantum computers, and can only embed small data sets (such as Iris and Bars and Stripes) reliably.
It is expected that future quantum computers would be larger and more reliable, making it possible to embed larger data sets.

Section \ref{sec:related} and \ref{sec:notation} cover the related work and notation used in this paper.
The quantum discriminator, associated quantum-classical training algorithm, theoretical analysis and notes on generalizability are presented in Section \ref{sec:quantum-discriminator}.
In Section \ref{sec:2-bit-classification} and Appendix \ref{sec:2-bit}, we apply the quantum discriminator to the 2-bit binary classification problem as a proof of concept.
In Section \ref{sec:empirical}, we benchmark the performance of our model on the Iris and Bars and Stripes data sets.

\section{Related Work}
\label{sec:related}
Several machine learning approaches on universal quantum computers have been proposed in the literature \cite{biamonte2017quantum}.
Lloyd and Weedbrook as well as Dallaire-Demers and Killoran derive the theoretical underpinnings of quantum generative adversarial networks \cite{lloyd2018quantum,dallaire2018quantum}.
Blance and Spannowsky propose a variational quantum classifier for use in high energy physics applications \cite{blance2021quantum}.
Shingu et al. propose a variational quantum algorithm for Boltzmann machine learning \cite{shingu2021boltzmann}.
Benedetti et al. propose quantum parameterized circuits as machine learning models.
Quiroga et al. propose a quantum k-means (QK-Means) clustering technique to discriminate quantum states on the IBM Bogota quantum device \cite{quiroga2021discriminating}.
Apart from universal quantum computing approaches, adiabatic quantum machine learning approaches have also been proposed for traditional machine learning models such as regression and k-means clustering \cite{date2021adiabatic,arthur2021balanced}.

The above literature describes quantum implementations of conventional machine learning models. 
Another line of research in quantum machine learning focuses on developing purely quantum or hybrid quantum-classical models that are novel and different from conventional machine learning models.
Gambs recasts the quantum discrimination problem within the framework of machine learning and uses the notion of learning reduction to solve different variants of the quantum classification task \cite{gambs2008quantum}.
Sentis et al. present a quantum learning machine for binary classification of qubit states that does not require quantum memory and produces classifiers that are robust to an arbitrary amount of noise \cite{sentis2012quantum}.
Chen et al. propose a discrimination method for two similar quantum systems and apply it to quantum ensemble classification \cite{chen2016quantum}.
Sergioli et al. propose a new quantum classifier called the Helstrom Quantum Centroid, which acts on density matrices that encode the classical patterns of a data set onto the quantum computer \cite{sergioli2019new}.
Park et al. focus on the squared overlap between quantum states as a similarity measure and examine the essential ingredients for the quantum binary classification, advancing the theory of quantum kernel-based binary classification \cite{park2020theory}.
In order to reduce the number of trainable parameters of a quantum circuit, Li et al. propose the Variational Shadow Quantum Learning (VSQL) framework \cite{li2020vsql}.
Blank et al. present a distance-based quantum classifier whose kernel is based on the quantum state fidelity between training and test data \cite{blank2020quantum}. 
They also conduct proof of principle experiments on the IBMQ platform.

A number of approaches to quantum machine learning leverage the two-step workflow followed by traditional machine learning models.
Havlicek et al. propose a quantum variational classifier and a quantum kernel estimator for classification problems \cite{havlivcek2019supervised}.
Schuld and Killoran propose a nonlinear feature map that maps data to a quantum feature space and discuss two discriminant models for classification \cite{schuld2019quantum}.
Bergou and Hillery propose a quantum discriminator that can distinguish between two unknown quantum states \cite{bergou2005universal}.
Lloyd et al. present a two-part quantum machine learning model, where the first part of the circuit implements a quantum feature map that encodes classical inputs into quantum states, and the second part of the circuit executes a quantum measurement, which acts as the output of the model. \cite{lloyd2020quantum}

Both universal as well as adiabatic approaches for quantum machine learning have been proposed in the literature.
However, there are several limitations in the current state-of-the-art.
Most of the proposed approaches are based on variational quantum circuits.
By definition, these are iterative quantum-classical hybrid approaches, and require multiple data exchanges between the quantum and the classical computer. 
These data exchanges necessitate frequent measurements of the quantum states, which introduce measurement errors as well as collapse the quantum superposition.
When the superposition is collapsed frequently, the ability of quantum computers to operate efficiently in high-dimensional tensor product spaces is curtailed.
So, it is desirable to minimize the use of classical computers in quantum machine learning methods.
With regards to validating quantum machine learning methods, most of the approaches in the literature are not validated on benchmark data sets such as Iris and MNIST---this is largely due to the small size and noisy nature of today's quantum computers.


The quantum discriminator proposed in this paper mitigates some of these challenges.
Firstly, it uses classical computers to process $\mathcal{O}(N)$ data during training only.
The classical computer is not used in the inferencing stage.
This restricted use of the classical computer enables the quantum discriminator to operate in the high-dimensional tensor-product spaces efficiently.
Next, we validate the quantum discriminator on the Iris and the Bars and Stripes data sets, which, in quantum machine learning, are some of the largest data sets that can be embedded on today's quantum computers.
Ideally, we would like to validate the quantum discriminator on the MNIST, CIFAR, ImageNet and other benchmark machine learning data sets.
However, the small size and low reliability of current quantum computers prevent us from embedding these large data sets onto them.
Another advantage of the quantum discriminator is that it can be trained in $\mathcal{O}(N \log N)$ time using $\mathcal{O}(N \log N)$ classical bits and $\mathcal{O}(\log N)$ qubits.
Inferencing can be performed on the quantum discriminator in $\mathcal{O}(N)$ time using $\mathcal{O}(\log N)$ qubits.


\section{Notation}
\label{sec:notation}

We use the following notation throughout this paper:
\begin{itemize}
    \item $\mathbb{R}$, $\mathbb{N}$, $\mathbb{B}$: Set of real numbers, natural numbers and binary numbers ($\mathbb{B} = \{0,1\}$) respectively.
    \item $N$: Number of data points in training data set ($N \in \mathbb{N}$).
    \item $d$: Dimension of each data point in the training data set ($d \in \mathbb{N}$).
    \item $b$: Dimension of each data point in the binary feature set of the training data set ($b \in \mathbb{N}$). 
    \item $B$: Number of unique states that can be attained using $b$ bits ($B = 2^b$).
    \item $X$: The training data set ($X \in \mathbb{R}^{N \times d}$).
    \item $Y$: The training labels ($Y \in \mathbb{B}^N$). If the $i^{\text{th}}$ data point belongs to class $0$ (class $1$), then $y_i = 0$ ($y_i = 1$).
    \item $\hat{X}$: The binary feature set of the training data set $X$ ($\hat{X} \in \mathbb{B}^{N \times b}$). $\hat{x}_i \in \hat{X}$ contains the features corresponding to the $i^{\text{th}}$ data point $x_i \in X$.
    \item $P$: The labels predicted by the quantum binary classification model ($P \in \mathbb{B}^{N}$). Ideally, the predicted labels should be identical to the training labels ($Y$).
\end{itemize}

\section{The Quantum Discriminator}
\label{sec:quantum-discriminator}

\begin{figure}[t!]
    \centering
    \includegraphics[trim={35px 125px 10px 200px}, clip, scale=0.25]{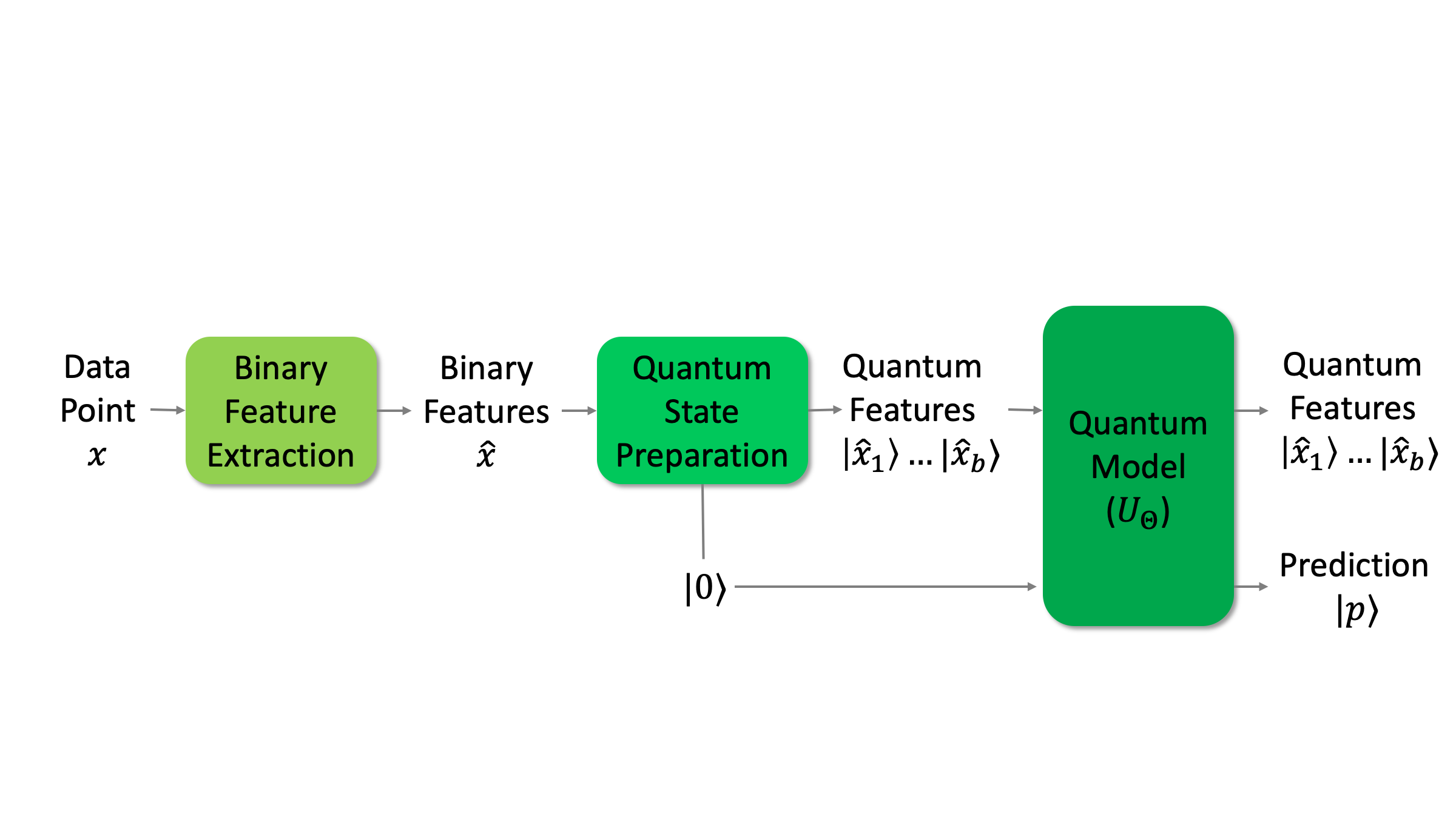}
    \caption{Model workflow.}
    \label{fig:model}
\end{figure}


Given a data point $x$ which belongs to one of two classes (Class $0$ or Class $1$), we would like to predict the correct class for $x$.
Generally, the binary classification model is characterized by a set of model parameters $\Theta$.
The workflow used for traditional machine learning classification models---such as support vector machines (SVM) and logistic regression---is comprised of 2-steps: (i) Feature extraction from the data; and (ii) Class determination by application of a discriminant function.
The workflow governing our quantum model for binary classification is analogous to this 2-step workflow as shown in Figure \ref{fig:model}.

In Figure \ref{fig:model}, we are given a data point $x \in \mathbb{R}^d$ whose class needs to be predicted.
We first extract the binary features of $x$, denoted by $\hat{x} \in \mathbb{B}^b$.
Usually, the extracted features are domain specific, for example, Histogram of Oriented Gradients (HOG) \cite{shu2011histogram} and Scale-Invariant Feature Transform (SIFT) \cite{nguyen2014object} have been widely used in computer vision.
The feature space could also originate from dimensionality reduction techniques like Principal Component Analysis (PCA). 
So, we do not make any assumptions on the extracted features, except that they are binary. 
This feature set of the training data is denoted by $\hat{X}$ in Figure \ref{fig:model}.
Each point $\hat{x} \in \hat{X}$ is a point on a $b$-dimensional unit hypercube, and there are $B$ such points.
Given $\hat{x}$, its equivalent quantum feature state is denoted by $\ket{\hat{x}}$, which is a point in $b$-dimensional Hilbert space.

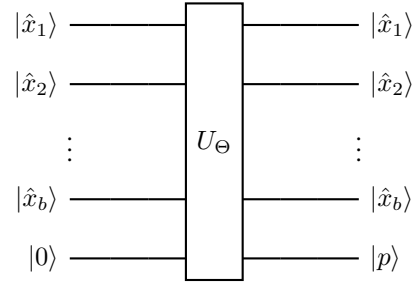
\begin{figure}[t!]
    \centering
    \begin{tikzpicture}
            \node[scale=1.0] {
            \begin{quantikz}
                \lstick{\ket{\hat{x}_{1}}}   & \qw & \qw & \gate[5, nwires={2,3,4,5}]{U_{\Theta}}  & \qw & \qw & \rstick{\ket{\hat{x}_{1}}} \qw \\
                \lstick{\ket{\hat{x}_{2}}}   & \qw & \qw & \qw                   & \qw & \qw & \rstick{\ket{\hat{x}_{2}}} \qw \\
                \vdots                  &     &     &                       &     &     & \vdots \\
                \lstick{\ket{\hat{x}_{b}}}   & \qw & \qw & \qw                   & \qw & \qw & \rstick{\ket{\hat{x}_{b}}} \qw \\ \lstick{\ket{0}}        & \qw & \qw & \qw                   & \qw & \qw & \rstick{\ket{p}} \qw
            \end{quantikz}
            };
            \end{tikzpicture}
    \caption{The quantum discriminator.}
    \label{fig:quantum-model}
\end{figure}

In addition to preparing the quantum feature state in Figure \ref{fig:model}, we also prepare a qubit in the $\ket{0}$ state, which would serve as our prediction qubit $\ket{p}$.
The quantum feature state $\ket{\hat{x}} = \ket{\hat{x}_1 \ldots \hat{x}_b}$, as well as the prediction qubit in the $\ket{0}$ state serve as inputs to the quantum discriminator as shown in Figure \ref{fig:quantum-model}.
The quantum discriminator ($U_\Theta$) is a $2B \times 2B$ matrix, which takes as input $\ket{\hat{x}}$ and $\ket{0}$, and outputs $\ket{\hat{x}}$ and the prediction $\ket{p}$.
We now describe $U_\Theta$, which is parameterized by $\Theta = \{\theta_1, \theta_2, \ldots, \theta_B\}$, $\theta_i \in \mathbb{B}$, $\forall i = 1, 2, \ldots, B$.

\begin{align}
    \scriptsize{U_\Theta =  \begin{bmatrix}
                1 - \theta_1    & \theta_1      & 0             & 0             & \ldots    & 0             & 0 \\
                \theta_1        & 1 - \theta_1  & 0             & 0             & \ldots    & 0             & 0 \\
                0               & 0             & 1 - \theta_2  & \theta_2      & \ldots    & 0             & 0 \\
                0               & 0             & \theta_2      & 1 - \theta_2  & \ldots    & 0             & 0 \\
                \vdots          & \vdots        & \vdots        & \vdots        & \ddots    & \vdots        & \vdots \\
                0               & 0             & 0             & 0             & \ldots    & 1 - \theta_B  & \theta_B \\
                0               & 0             & 0             & 0             & \ldots    & \theta_B      & 1 - \theta_B
                \end{bmatrix}} \label{eq:u-sub-theta}
\end{align}

Since all quantum operators are unitary matrices, it is crucial that $U_\Theta$ be a unitary matrix.
We now show that $U_\Theta$ is unitary by showing that $U_\Theta^\dagger U_\Theta = U_\Theta U_\Theta^\dagger = I$.
Since $U_\Theta$ is symmetric, $U_\Theta^\dagger = U_\Theta$.
Because $\theta_i \in \mathbb{B}$, the off-diagonal elements in $U_\Theta^\dagger U_\Theta$ and $U_\Theta U_\Theta^\dagger$ are zeros.
The diagonal elements of $U_\Theta^\dagger U_\Theta$ and $U_\Theta U_\Theta^\dagger$ are of the form $(1 - \theta_i)^2 + \theta_i^2$, which always equals unity.
So, $U_\Theta^\dagger U_\Theta = U_\Theta U_\Theta^\dagger = I$. 
Thus, $U_\Theta$ is a unitary.


\subsection{Number of binary features}
\label{sub:num-features}

We would like to extract $b$ binary features so that the $N$ points in the feature set span as much of the feature space as possible.
This would ensure that the quantum discriminator trained as a result would be generalizable to any test data point that originates from the same distribution as the training data set.
Since the size of the feature space is $2^b$, we have: $N \approx 2^b$, or $b$ is $\mathcal{O}(\log N)$.

\subsection{Training the Quantum discriminator}
\label{sub:training}

\begin{figure}[t!]
    \centering
    \includegraphics[trim={20px 130px 75px 130px}, clip, scale=0.27]{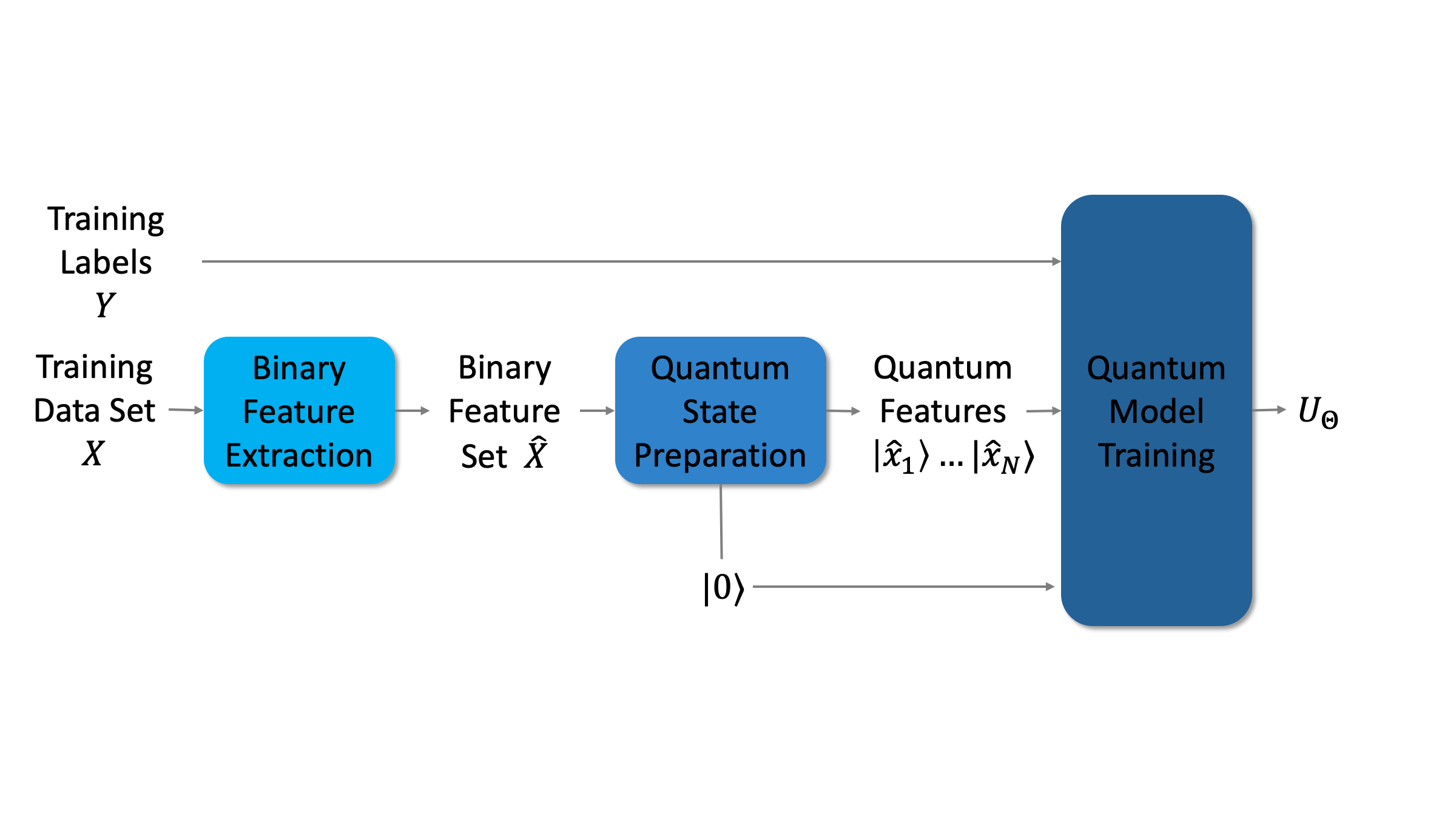}
    \caption{Training workflow.}
    \label{fig:training-workflow}
\end{figure}

Figure \ref{fig:training-workflow} shows the training workflow for our quantum model for classification.
We are given the training data set $X$ and the training labels $Y$.
We compute the binary feature set $\hat{X}$ and prepare the quantum feature states $\ket{\hat{x}_1} \ldots \ket{\hat{x}_N}$, where $\hat{x}_i \in \hat{X}, \quad \forall i = 1, 2, \ldots, N$.
In addition to the quantum feature states, we also prepare the $\ket{0}$ state.
Given $\ket{\hat{x}_1} \ldots \ket{\hat{x}_N}$, $Y$, and $\ket{0}$, the goal of training the quantum discriminator is to find the model parameters $\Theta$, that minimize a well-defined error function.

Some examples of error functions used in classification tasks are the Euclidean error (or $L2$-norm) in linear regression, negative log likelihood error in logistic regression, and the cross entropy error in deep neural networks.
Negative log likelihood and cross entropy errors are generally used in probabilistic machine learning models, whereas, $L2$-norm is used for non-probabilistic machine learning models.
For non-probabilistic machine learning models, the $L1$-norm is considered to be more robust than the $L2$-norm \cite{li2015robust}.
However, $L2$-norm is generally used in practice because it is differentiable and amenable to gradient computations.
Since our quantum discriminator is not probabilistic and does not require gradient computations, we will use the $L1$-norm as the error function.
Thus, the training process can be stated as follows:
\begin{align}
    \min_{\Theta} \ E(\Theta) = \frac{1}{N} \sum_{i=1}^N |p_i - y_i| \label{eq:error}
\end{align}

where $p_i$ is the measured value of $\ket{p_i}$ and the output state, $\ket{\hat{x}_i \ p_i}$ equals $U_\Theta \ket{\hat{x}_i \ 0} \ \forall i = 1, 2, \ldots, N$.



\begin{algorithm}
\caption{Training the quantum discriminator.}
\label{algo:training}
\begin{algorithmic}
    \STATE \textbf{Input}: $\hat{x}_1, \hat{x}_2, \ldots, \hat{x}_N, y_1, y_2, \ldots, y_N$
    \STATE $b \gets$ \texttt{length}($\hat{x}_1$)
    \STATE $B \gets 2^b$
    \STATE $\tau \gets [2^{b-1}, 2^{b-2}, \ldots, 2^{1}, 2^{0}]^T$
    \STATE $\theta_j \gets 0 \quad \forall j = 1, 2, \ldots, B$
    \STATE\algorithmicfor{$\quad i = 1, 2, \ldots, N$}
        \STATE $\quad$ \algorithmicif {$\quad y_i == 1$}
        \STATE $\quad\quad$ $j \gets 1 + \tau \cdot \hat{x}_i$
        \STATE $\quad\quad$ $\theta_j \gets 1$
\end{algorithmic}
\end{algorithm}

Algorithm \ref{algo:training} outlines the training process for the quantum discriminator.
The inputs to the model are the feature vectors $\hat{x}_1, \hat{x}_2, \ldots \hat{x}_N$, and the training labels $y_1, y_2, \ldots, y_N$.
We first initialize $b$ and $B$.
The $\texttt{length}(z)$ function computes the length of $z$.
We then set the vector $\tau = [2^{b-1}, 2^{b-2}, \ldots, 2^0]$. 
Next, we setup the quantum circuit shown in Figure \ref{fig:quantum-model}, where $U_\Theta = I$, by initializing all the model parameters $\theta_j$ to zero ($j = 1, 2, \ldots, B$).
We then look at each feature vector $\hat{x}_i$ ($i = 1, 2, \ldots, N$).
If $\hat{x}_i$ belongs to Class $1$ (i.e. $y_i = 1$), then we compute the index $j$ as $1 + \tau \cdot \hat{x}_i$, and set $\theta_j = 1$.
We repeat this process for all $N$ points in the training feature set $\hat{X}$. 
When Algorithm \ref{algo:training} terminates, it assigns all points in $\hat{X}$ to their respective correct classes.

We now shed some light on why Algorithm \ref{algo:training} works.
The input state to the quantum discriminator, $\ket{\hat{x} \ 0}$, exists in $(b+1)$-dimensional Hilbert space and is in a superposition of all $2B$ possible states.
As such, each of the $B$ quantum feature states $\ket{\hat{x}}$, occurs twice: as $\ket{\hat{x} \ 0}$ and $\ket{\hat{x} \ 1}$.
These two states can be interpreted as $\ket{\hat{x}}$ belonging to Class $0$ or Class $1$ respectively.
While training the quantum discriminator, we select the correct class for each $\ket{\hat{x}}$.
The rows and columns of $U_\Theta$ that correspond to $\hat{x}$ can be found at indices $j = 1 + \tau \cdot \hat{x}$ and $j+1$, which can be leveraged to assign $\hat{x}$ to Class $0$ or Class $1$ respectively.
Initially, the $2 \times 2$ sub-matrix at $j^{\text{th}}$ row and $j^{\text{th}}$ column of $U_\Theta$ is an identity matrix because $\theta_j$ is initialized to $0$.
If $\hat{x}$ belongs to Class $0$, then this sub-matrix outputs $\ket{\hat{x} \ 0}$ for the input $\ket{\hat{x} \ 0}$, which can be interpreted as $\hat{x}$ being assigned to Class $0$.
On the other hand, if $\hat{x}$ belongs to Class $1$, then this sub-matrix must be changed to the Pauli-$X$ gate (also called the bit-flip gate or the NOT gate), which is done by setting $\theta_j$ to $1$.
The Pauli-$X$ gate at $j^{\text{th}}$ row and $j^{\text{th}}$ column of $U_\Theta$ outputs $\ket{\hat{x} \ 1}$ for the input $\ket{\hat{x} \ 0}$, which can be interpreted as $\hat{x}$ being assigned to Class $1$.

\subsection{Theoretical Analysis}
\label{sub:theoretical-analysis}

We analyze the time and space complexity of training the quantum discriminator here.
In Algorithm \ref{algo:training}, lines $1$ and $2$ require $\mathcal{O}(1)$ time and line $3$ requires $\mathcal{O}(b)$ time.
It may seem that line $4$ requires $\mathcal{O}(B)$ time, but initializing $\theta_j$ to $0$ essentially refers to setting up the quantum circuit with $U_\Theta = I$.
This entails setting up $b+1$ qubits, which takes $\mathcal{O}(b)$ time.
Computing the dot product on line $7$ may take $\mathcal{O}(b)$ time and setting $\theta_j$ to unity on line $8$ takes $\mathcal{O}(1)$ time.
Since we may repeat lines $7$ and $8$ $N$-times in the worst case, the time complexity of Algorithm \ref{algo:training} is $\mathcal{O}(N b)$, which is the same as the size of the feature set $\hat{X}$.
Since $b$ is $\mathcal{O}(\log N)$ from Section \ref{sub:num-features}, the time complexity is $\mathcal{O}(N \log N)$.
Since we use $\mathcal{O}(Nb)$ classical bits for storing $\hat{X}$, $Y$ and $\tau$, and computing the dot product on line $7$, the space complexity of Algorithm \ref{algo:training} is also $\mathcal{O}(N \log N)$.
The qubit footprint of Algorithm \ref{algo:training} is $\mathcal{O}(b)$ because we use $b+1$ qubits.
Thus, it is possible to train the quantum discriminator shown in Figure \ref{fig:quantum-model} in $\mathcal{O}(N \log N)$ time, using $\mathcal{O}(N \log N)$ classical bits and $\mathcal{O}(b)$ qubits. Subsequently, inferencing can be performed on a given input using the quantum circuit seen in Figure \ref{fig:quantum-model} in $\mathcal{O}(N)$ time using just $\mathcal{O}(b)$ qubits.

\subsection{Generalizability}
\label{sub:generalizability}

By generalizability, we refer to the ability of a machine learning model to make predictions on data points not encountered during training.
The quantum discriminator has the ability to classify points in the training data set with $100\%$ accuracy owing to an exponential number of model parameters ($\theta_1, \theta_2, \ldots, \theta_B$). 
It is highly complex and highly susceptible to overfitting the training data.
This affinity to overfit is kept in check by the feature extraction process.
If the extracted  binary features are good and small in number ($b \approx \mathcal{O}(\log N)$), the number of model parameters are also small ($B \approx \mathcal{O}(N)$).
The subsequent quantum discriminator would have a lower tendency of overfitting and would be generalizable.

\section{2-Bit Binary Classification}
\label{sec:2-bit-classification}
As a proof of concept, we use the quantum discriminator to solve the 2-bit binary classification problem.
Because of the large number of figures in this part of the study, we have explained it in Appendix \ref{sec:2-bit}.
The reader is advised to read Appendix \ref{sec:2-bit} to gain a better understanding of the quantum discriminator and its application.

\section{Empirical Evaluation}
\label{sec:empirical}

\subsection{Hardware and Simulator Details}
\label{subsec:hardware-and-simulator}

We validated the performance of the quantum discriminator on the Iris and Bars and Stripes data sets.
All our experiments were conducted on the IBM Jakarta processor as well as in numerous noise-free simulations using the QASM simulator in IBM Qiskit. The IBM Jakarta quantum computer consists of $7$ qubits with a total quantum volume of $16$. The average CNOT and readout errors on this machine were $0.015$ and $0.035$ respectively.
The classical computer used in the training process was a desktop workstation having Intel Core i4-4670K CPU, running at 3.4 GHz, 64-bit operating system and 16 GB RAM.

\subsection{The Iris data set}
\label{subsec:iris}

The Iris data set is pervasive as a benchmark in machine learning, and is small enough to be embedded on current quantum computers. 
The data set contains $150$ data points gathered from a sample of Iris flowers. Each data point consists of four measurements taken on the given Iris flower along with its species. The recorded attributes for each flower are petal length, petal width, sepal length, and sepal width measured in centimeters. The $150$ data points are split evenly between three species of Iris: {\it{Iris-Setosa}},{\it{ Iris-Versicolor}}, and {\it{Iris-Virginica}}. 

Using this data set, a model can be created whereby the petal/sepal lengths and widths of a given Iris datum can be used to predict its species. For the purpose of testing the quantum discriminator---which is designed for \textit{binary} classification---the data set was restricted to just the {\it Iris-Setosa} and {\it Iris-Virginica} samples, which were labeled as Class 0 and Class 1 respectively.  
This reduced the size of the data set to $100$ data points.

\subsubsection{Feature extraction}
Before training the quantum discriminator, binary features must be extracted from the data.  
In this case, three features were gathered from the data by imposing threshold values for the sepal length, sepal width, and petal length. 
The fourth attribute, petal width, was not considered in this feature extraction process. 
Specifically, our extracted data consisted $3$-tuples of binary numbers, $\hat x = (a_1, a_2, a_3) \in \mathbb{B}^3$, where $a_1 = 1$ if the sepal length was recorded to be above $5.50$ cm, $a_2 = 1$ if the sepal width was recorded to be above $3.00$ cm, $a_3 = 1$ if petal length was recorded to be above $3.00$ cm, and each $a_i$ was set to $0$ if it failed to meet these respective threshold values.  
This feature extraction procedure resulted in a separable data set. 
Moreover, the extracted data set was found to span only three-quarters of our entire binary feature space, which has a theoretical size of $2^3 = 8$, i.e. $8$ unique configurations of binary features.

\subsubsection{Training}
Training experiments were conducted in which the $100$ data points were partitioned at random into a training data set of size $N$, and the remaining $100-N$ data points were reserved for validation. 
This parameter $N$, was varied from $N=4$, where at most half of feature space could be sampled, to $N=80$, which corresponds to a train/test split scheme commonly used to evaluate machine learning approaches.  
Training was performed in accordance with Algorithm \ref{algo:training}.

The theory behind the quantum discriminator prescribes the number of features $(b)$ to be $\mathcal{O}(\log N)$.
Since the size of the smallest and largest training sets in our experiments were $4$ and $80$ respectively, we would need to have between $\log_2 4$ and $\log_2 80$ binary features, i.e., $2$--$6$ binary features.
As mentioned above, we selected $3$ binary features for this task, which is consistent with the assumptions of the quantum discriminator.


In each trial, a quantum circuit was constructed using IBM's Qiskit software development kit in order to evaluate our model on each point in the validation set. 
Specifically, for each point in validation, $(x_i, y)$, a circuit was constructed which initializes a blank quantum register of $4$ qubits into the state $| \hat x_i \ 0 \rangle$ before passing into a series of gates equivalent to the unitary transformation obtained from the model parameters which were extracted in training. The predicted label, $p$, was then recorded for comparison with its true value $y$.

\subsubsection{Results on the Iris data set} 

\begin{figure}[t!]
    \centering
    \includegraphics[clip, scale=.044]{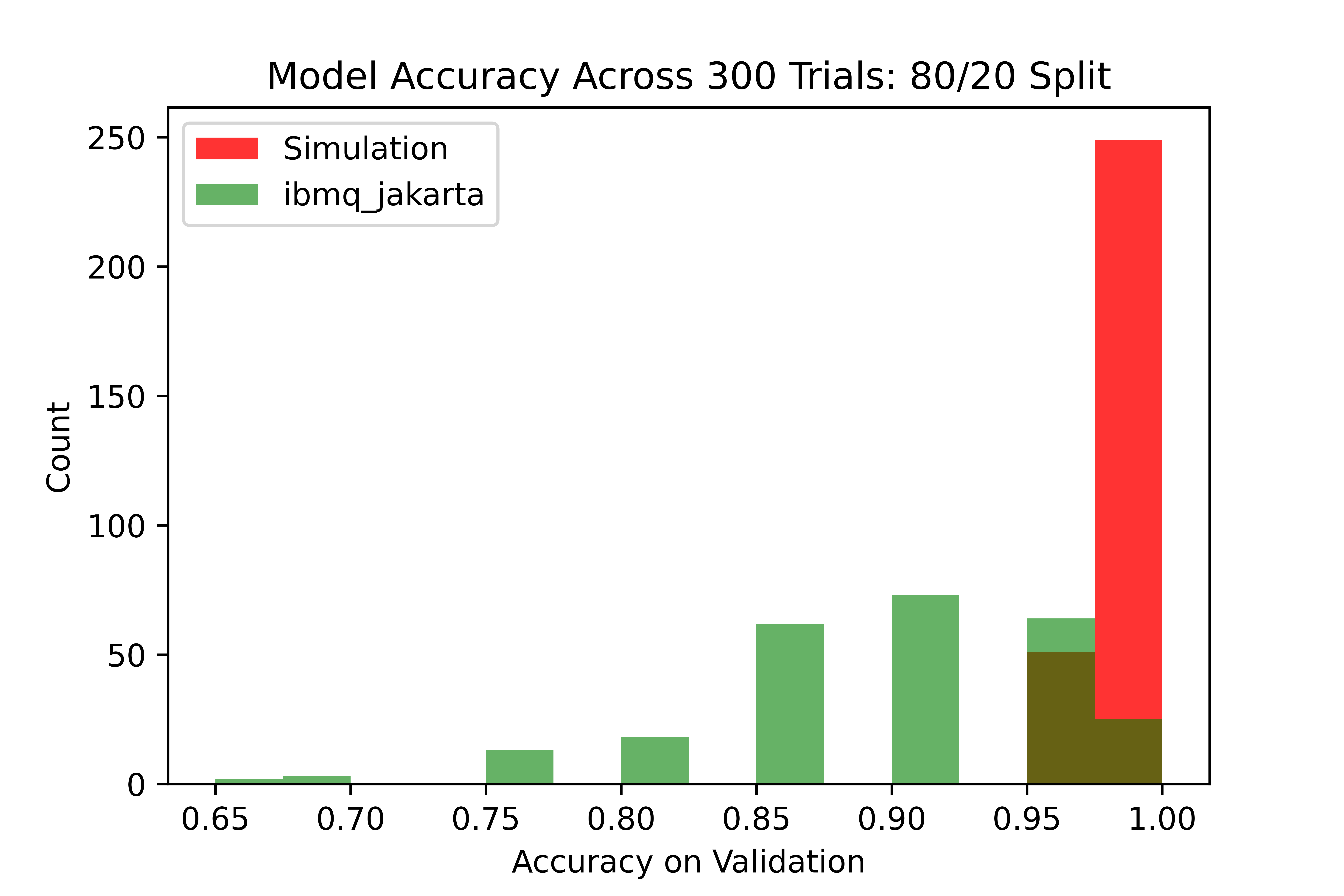}
 \caption{Histogram of validation accuracies on hardware and simulator in the case of $N=80$ for the Iris data set across $300$ trials each.}
    \label{fig:8020}
\end{figure}

 \begin{figure}[t!]
    \centering
    \includegraphics[clip, scale=.044]{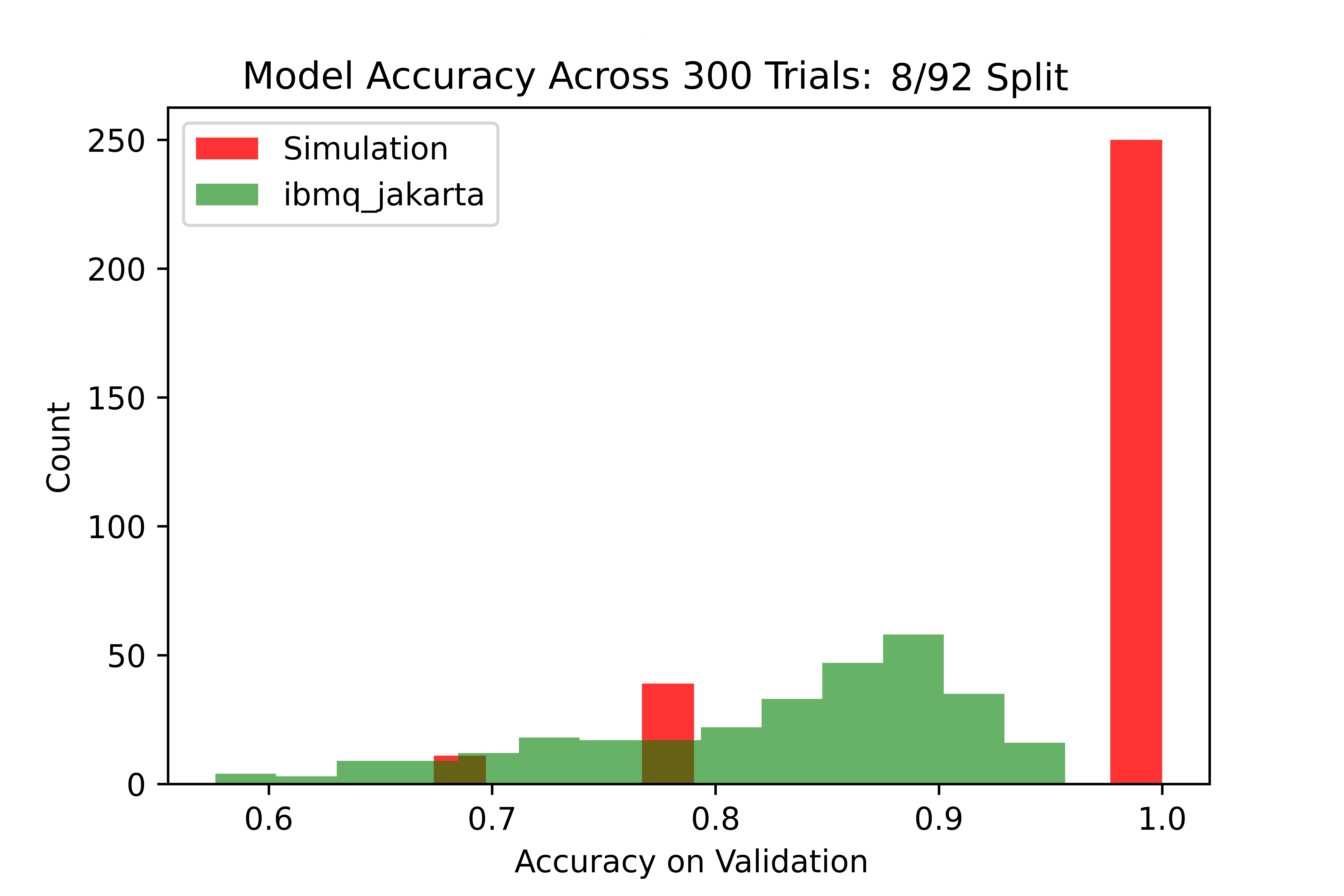}
 \caption{Histogram of validation accuracies on hardware and simulator in the case of $N=8$ for the Iris data set across $300$ trials each.}
    \label{fig:892}
\end{figure}

\begin{figure}[t!]
    \centering
    \includegraphics[clip, scale=.5]{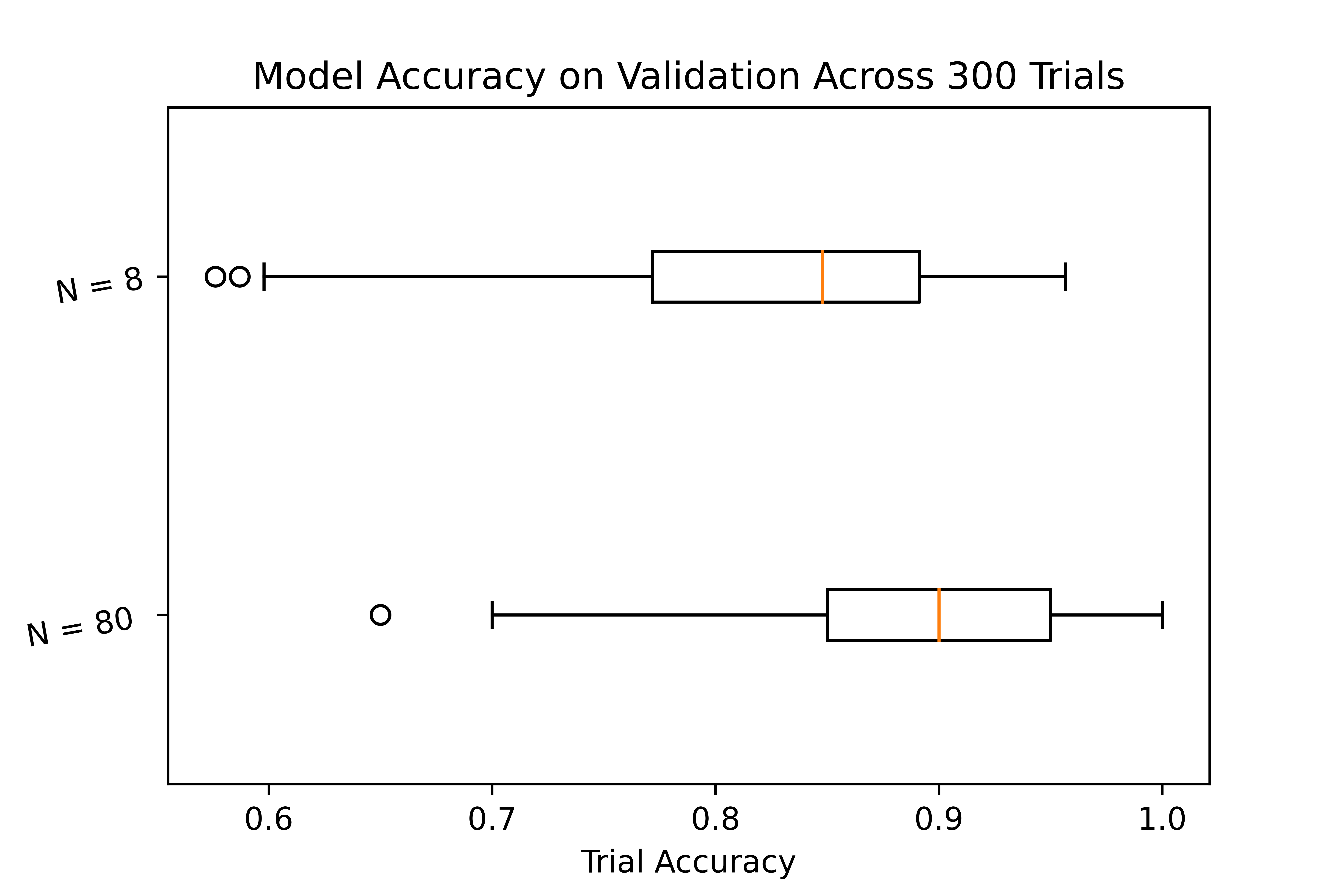}
 \caption{Box plot of model accuracies on the IBM Jakarta processor for $N=80$ and $N=8$ across $300$ trials each.}
    \label{fig:bw}
\end{figure}

The three training sizes tested in this work are $N= 80, 8,$ and $4$. In the former two cases, experiments consisting of $300$ trials each were conducted, with each experiment being performed once in the QASM simulator and once again on the IBM Jarkarta processor, meaning there were four experiments in total for these two cases. In each trial, $N$ data points were selected at random to form a training set, which was used to train the quantum discriminator in accordance with Algorithm \ref{algo:training}. 
The trained model was used for inferencing on each point in the validation set for benchmarking purposes. 
Benchmarking on the $N=4$ case was conducted analogously, except just one experiment was conducted (in simulation) and the number of trials was increased to $600$. 

In the case of $N=80$, the discriminator obtained an average validation accuracy of $99.15\%$ with a standard deviation of $1.878\%$ in simulation. On the Jakarta processor, however, the discriminator obtained a much lower accuracy of $89.13\%$ on average with a standard deviation of $6.97\%$. A histogram depicting the distribution of model accuracies across the $300$ trials in both cases is depicted in Figure \ref{fig:8020}.

When the training set was lowered to size $N=8$, the average validation accuracy dropped to $94.98\%$ with a standard deviation of $9.047\%$ in simulation, whereas the average accuracy on the Jakarta processor fell to $82.37\%$ with a standard deviation of $8.403\%$. The distribution of model accuracies on validation is similarly displayed in Figure \ref{fig:892}. 
Figure \ref{fig:bw} displays a box plot of model accuracies on quantum hardware for $N$ equals $80$ and $8$.


\begin{figure}[t!]
    \centering
    \includegraphics[clip, scale=.52]{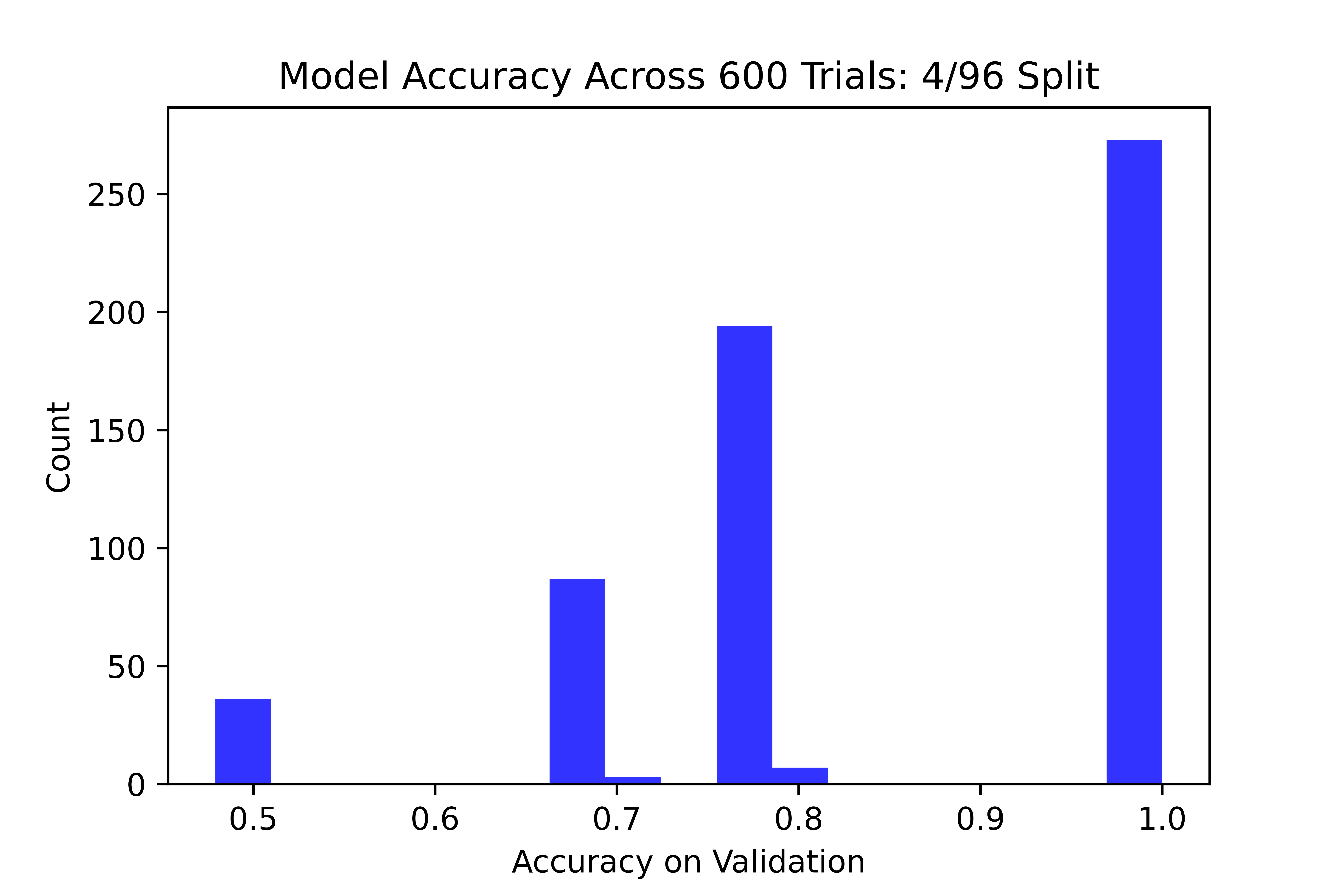}
 \caption{Histogram of validation accuracies on simulator in the case of $N=4$ for the Iris data set across $600$ trials.}
    \label{fig:496}
\end{figure}

\begin{figure}[t!]
    \centering
    \includegraphics[clip, scale=.5]{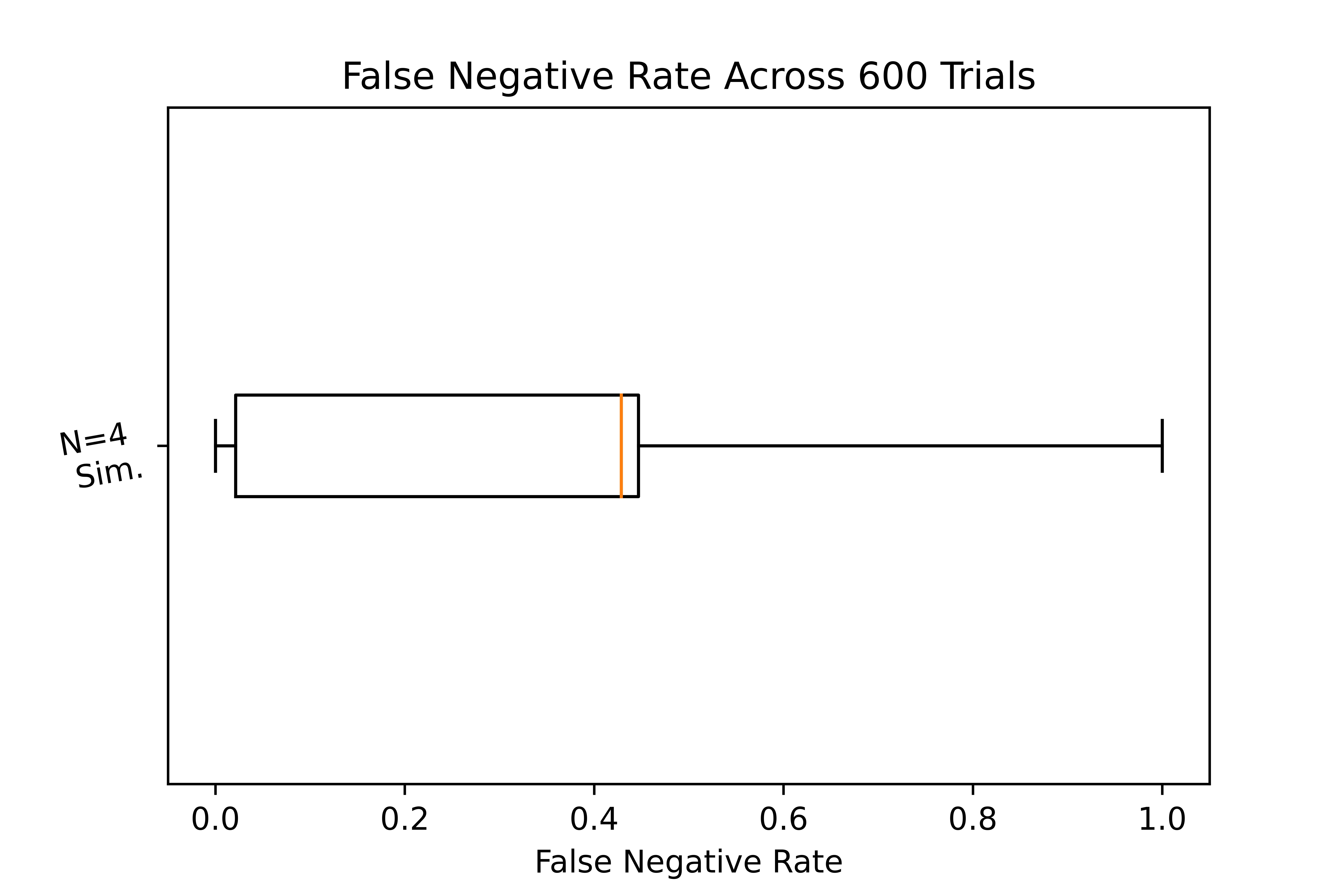}
 \caption{Box plot of model accuracies on simulator across $600$ trials on the IBM Jakarta processor for $N=4$}
    \label{fig:bw2}
\end{figure}

In the case of $N=4$, the average model accuracy across our $600$ trials fell to $84.41\%$, with an increased standard deviation of $15.04\%$; the distribution of which can be seen in Figure \ref{fig:496}. Additional statistics were gathered in this case. Inaccurate predictions made on validation were separated into Type I (false positive) and Type II (false negative) errors. It was found that all errors made by the models belong to Type II, meaning the model had a false positive rate of $0$ in each simulated trial.
In other words, each simulated model had a precision of $100\%$. It was found that false negative rate (also called the {\it miss rate}) was $0.3074$ on average with a standard deviation of $0.2934$. This distribution, seen in Figure \ref{fig:bw2} was heavily skew-right; i.e. the distribution was more concentrated on the side closer to a false-negative rate of $0$.  Consequently, the average recall (also called {\it sensitivity} or {\it hit rate}) was $0.6925$ on average with the same standard deviation. 

While these results are reported for the Iris data set and should be interpreted accordingly, we believe these are still very interesting.
The fact that the quantum discriminator can achieve $84\%$ accuracy when trained on just $4\%$ of the data can be used to build approximate machine learning models with moderately high accuracy very quickly.
Such models could then be refined quickly in subsequent training iterations, thereby reducing the training times.

\subsection{The Bars and Stripes Data Set}
\label{sec:bns}

\begin{figure}[t!]
\centering
\includegraphics[clip, scale=.11]{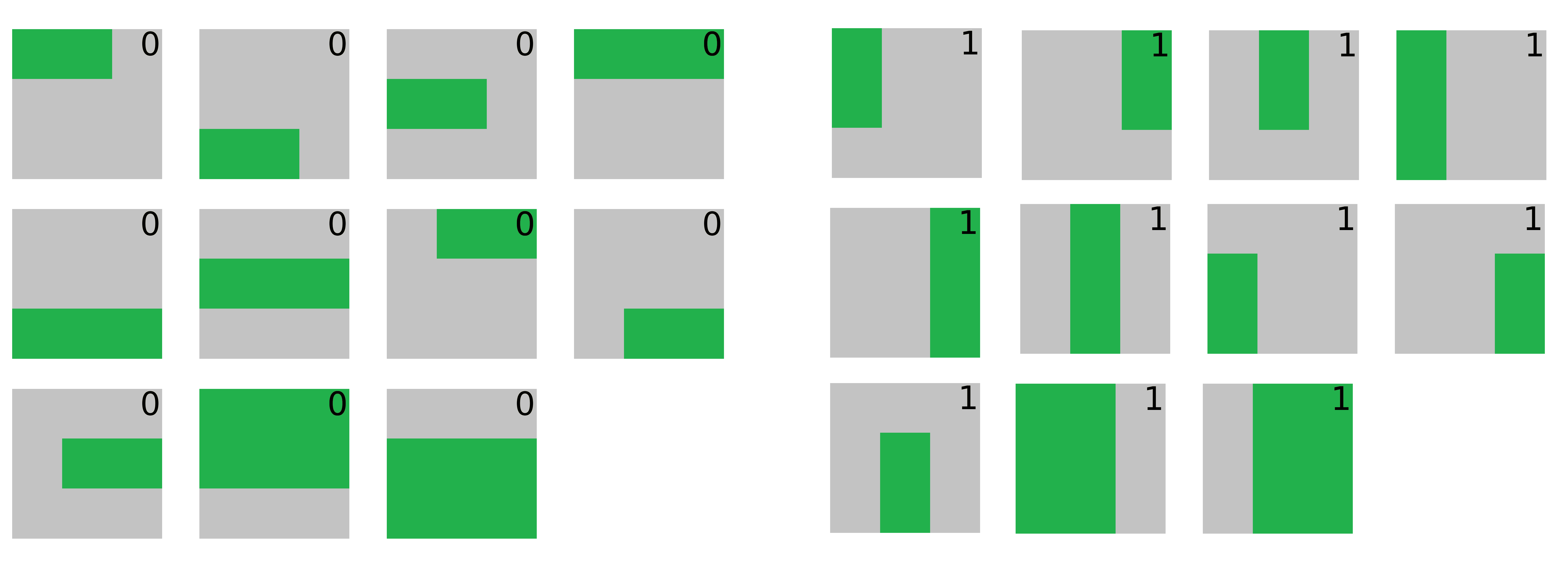}
\caption{A diagram depicting all possible bars (on the left) and stripes (on the right) in our data set. Here, cells are illuminated in green. Bars and stripes are assigned the classes $0$ and $1$ respectively.}
\label{fig:BAS}
\end{figure}

To further evaluate the performance of our quantum discriminator, a Bars and Stripes data set of size $100$ was generated on a $3$x$3$ grid. Each cell in this grid can either be illuminated or not. 
In this experiment, a grid is said to be a bar if it has a rectangle of illuminated cells whose width is strictly greater than its height.
Similarly, a grid is said to be a stripe if it has a rectangle of illuminated cells whose height is strictly greater than its width.
Accordingly, all bars and stripes viable under this formulation are displayed in Figure \ref{fig:BAS}. Using this scheme, $100$ samples were drawn from a uniform sample of these viable bars and stripes to form the data set used in this experiment. The bars and strips were assigned to class $0$ and $1$ respectively.
The Bars and Stripes experiments were run in simulation only.

\subsubsection{Methodology}
Nine binary features were extracted from each data point by reading the cells of each grid in lexicographic order, recording a one if the given cell is illuminated and a zero otherwise.
Two experiments were conducted on the Bars and Stripes data set in a similar fashion to the Iris data set in Section \ref{subsec:iris}. 
Specifically, $300$ trials (each) were conducted in noise-free simulation whereby the data set was randomly partitioned into a training set of size $N$ and the remaining data were reserved for validation of the resulting model; the training sizes used in this case were $N = 80$ and $N=11$. The latter quantity was chosen as it is the minimal number of points needed to sample all of the Class $1$ data in our binary feature space. 
    
\subsubsection{Results on the Bars and Stripes Data Set}
\begin{figure}[t!]
    \centering
    \includegraphics[clip, scale=.5]{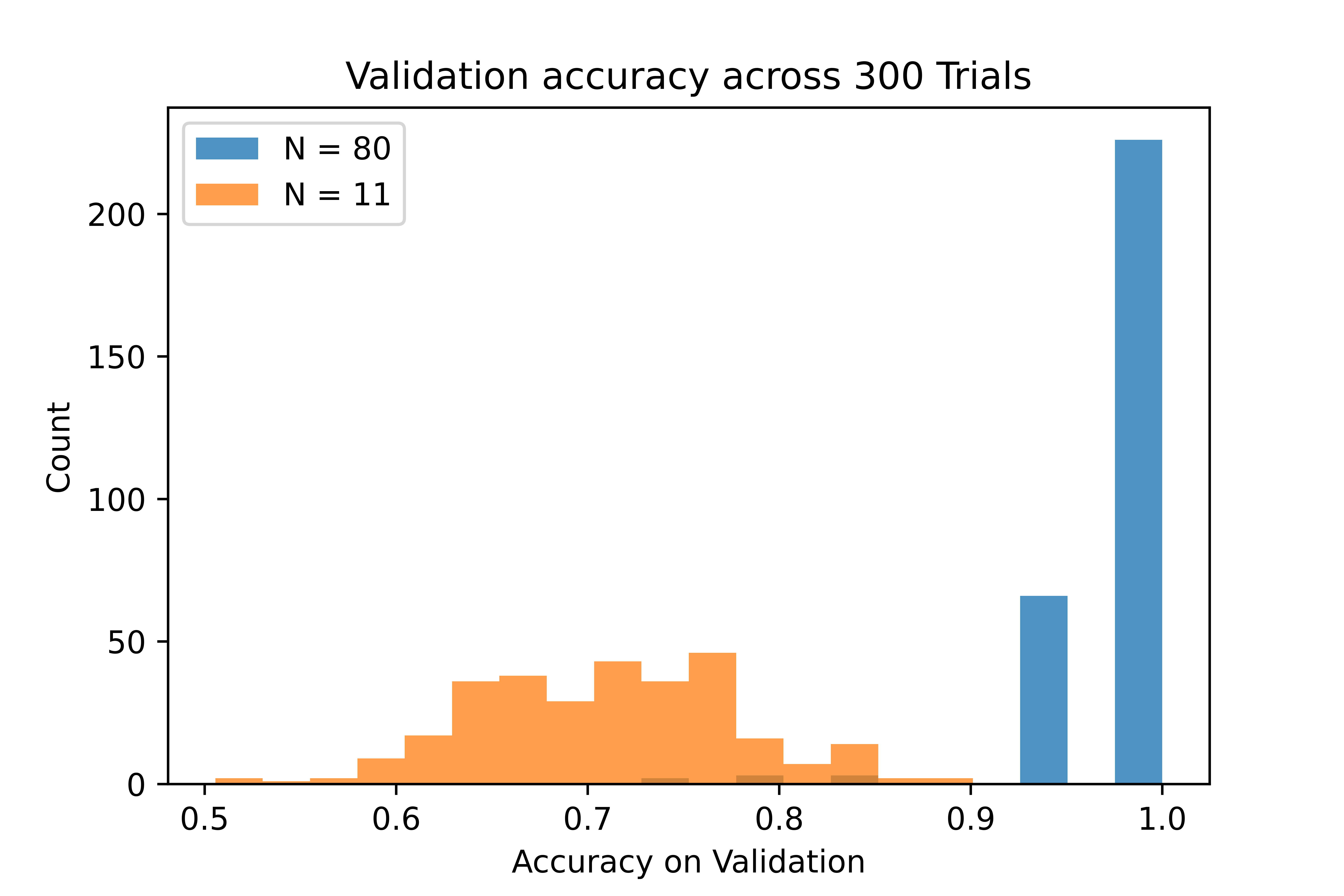}
 \caption{Histogram of validation accuracies on simulator in the case of $N=80$ and $8$ for the Bars and Stripes data set across $300$ trials each.}
    \label{fig:bas-his}
\end{figure}

In the case of $N=80$, the discriminator obtained an average validation accuracy of $98.38\%$ with a standard deviation of $3.647\%$ in simulation. When $N$ was lowered to $11$, the average validation accuracy dropped to $71.02\%$ with a standard deviation of $6.765\%$ in simulation. Figure \ref{fig:bas-his} displays how the distribution of model accuracies on validation varies in simulation with this change in $N$. 
These results were very similar to the Iris data set.
    
\subsection{Discussion}


The Quantum Discriminator fared extremely well in both of the cases where the size of the feature set was $\mathcal{O}(\log N)$ for the Iris as well as the Bars and Stripes data sets. 
There was a noticeable gap ($\approx 10 \%$) in both Iris cases ($N=80$ and $8$) between the discriminator's performance in simulation and on the IBM Jakarta processor. 
This discrepancy can be attributed to the fact that the state of the qubit as well as the inter-qubit connections on the quantum hardware are extremely sensitive to any kind of noise (thermal, radiation, vibrational etc.) in today's quantum computers.
Consequently, it is extremely difficult to maintain the qubit states for longer periods of time, and the qubits have a tendency to lose their state during the computation---this is called decoherence.
Qubit decoherence is known to result in considerably poor performance on the hardware as compared to simulation for many quantum algorithms.

To mitigate this issue of decoherence, error correction regimes are often incorporated into quantum algorithms.
However, no error-correction was performed in this experiment as it adds considerable overheads in terms of both time and space complexity.
It did not seem warranted considering that only four qubits were required for the Iris experiments, and only ten qubits were required for the Bars and Stripes experiments under our feature extraction regime. 
It stands to reason that the disparity in performance of the simulated and real-world models would increase dramatically as the number of qubits required for inferencing increases when working with today's quantum computers.
Having said that, it is expected that with hardware and engineering improvements in the future, quantum computers would become less noisy, more reliable, and larger in size.
These fault-tolerant quantum computers are expected to run quantum algorithms at par with (if not better than) the current simulation results.
In both data sets, it was observed that the quantum discriminator can obtain moderately high accuracies even when the training set is sparse, meaning approximate models can be quickly and efficiently generated for subsequent refinement. 


\section{Conclusion}
\label{sec:conclusion}


Alternative computing paradigms, such as quantum computing, present a new frontier in which novel machine learning techniques can be developed to address the current limitations of classical learning techniques. In this work, we outline a quantum machine learning technique for binary classification called the quantum discriminator.
The quantum discriminator is used as a discriminant function to infer the label of a given datum from its extracted features. 
The quantum discriminator is a $2B \times 2B$ unitary matrix, which is parameterized by $B$ parameters. 
It can be trained in $\mathcal{O}(N \log N)$ time, using $\mathcal{O}(N \log N)$ classical bits and $\mathcal{O}(b)$ qubits.
Inferencing on the quantum discriminator can be performed in $\mathcal{O}(N)$ time using $\mathcal{O}(b)$ qubits.
We demonstrated that this model can be used to completely solve the 2-bit binary classification problem as outlined in Appendix \ref{sec:2-bit}.
Furthermore, we evaluated its performance on the Iris data set, which is a benchmark machine learning data set, and also on a $3$x$3$ Bars and Stripes data set.
This empirical evaluation demonstrates the discriminator's potential to generate highly accurate, inherently precise models on a separable data sets when appropriate binary features have been extracted from the data. 

In our future work, we would like to evaluate the discriminator's performance on problems involving larger, more complex data sets, such as MNIST.
We would also like to extend the quantum discriminator to multi-class classification problems.
Lastly, we would like to investigate purely quantum training algorithms for training the quantum discriminator as opposed to the hybrid quantum-classical training algorithm described in this paper.

\bibliography{example_paper}
\bibliographystyle{icml2022}

\newpage
\appendix
\onecolumn
\section{Appendix: 2-bit Binary Classification as a Proof of Concept}
\label{sec:2-bit}

\begin{figure}[h!]
    \centering
    \subfigure[Case 1]{
        \includegraphics[scale=0.36,trim={70px 55px 70px 55px}, clip]{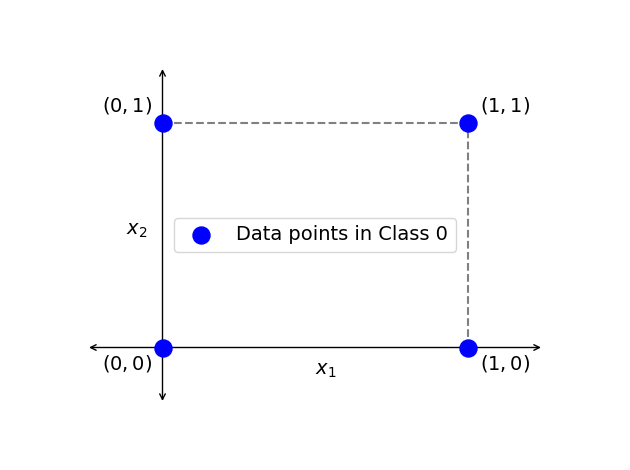}
    }
    \subfigure[Case 2]{
        \includegraphics[scale=0.36,trim={70px 55px 70px 55px}, clip]{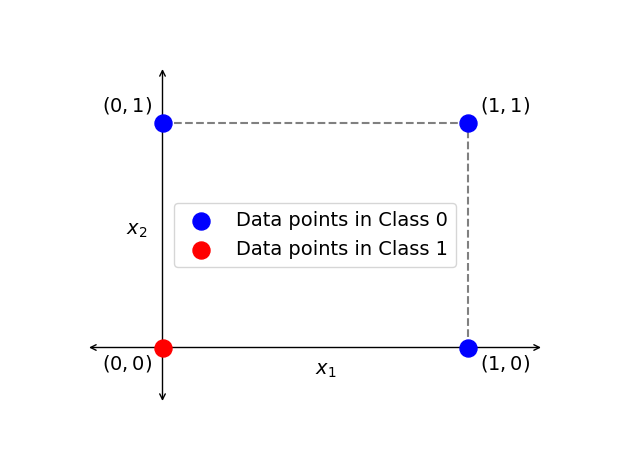}
    }
    \subfigure[Case 3]{
        \includegraphics[scale=0.36,trim={70px 55px 70px 55px}, clip]{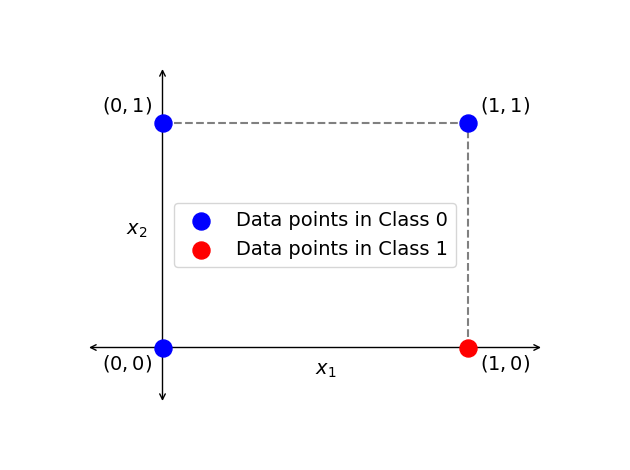}
    }
    \subfigure[Case 4]{
        \includegraphics[scale=0.36,trim={70px 55px 70px 55px}, clip]{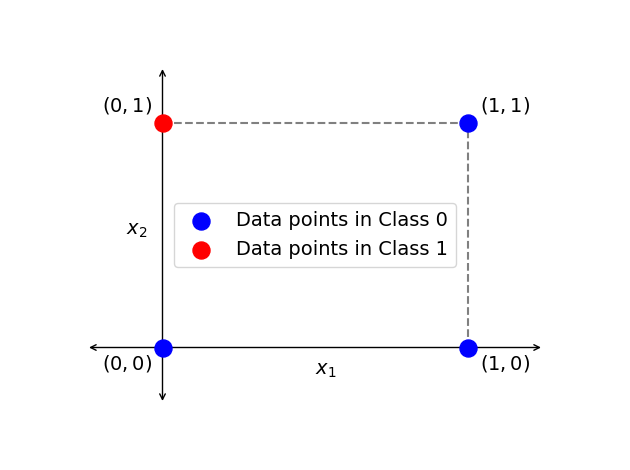}
    }
    \subfigure[Case 5]{
        \includegraphics[scale=0.35,trim={70px 55px 70px 55px}, clip]{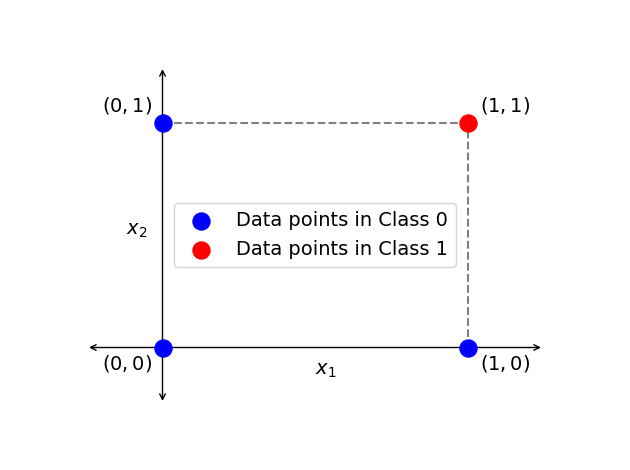}
    }
    \subfigure[Case 6]{
        \includegraphics[scale=0.35,trim={70px 55px 70px 55px}, clip]{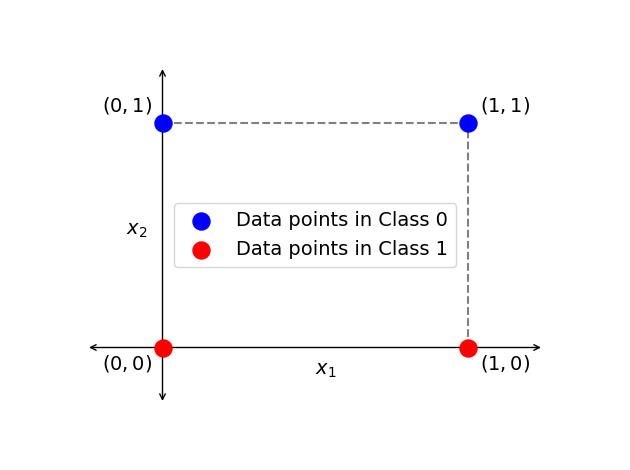}
    }
    \subfigure[Case 7]{
        \includegraphics[scale=0.35,trim={70px 55px 70px 55px}, clip]{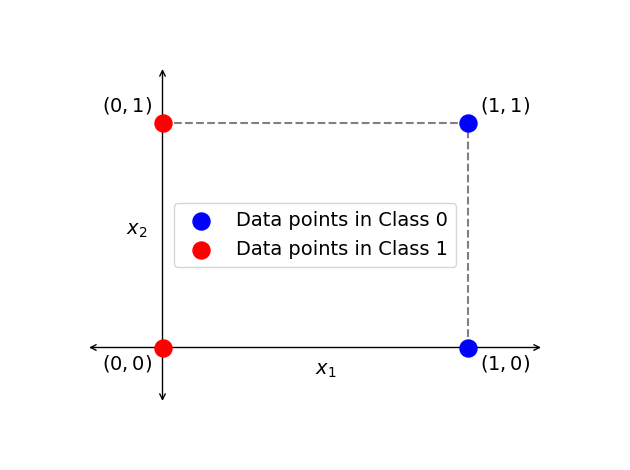}
    }
    \subfigure[Case 8]{
        \includegraphics[scale=0.35,trim={70px 55px 70px 55px}, clip]{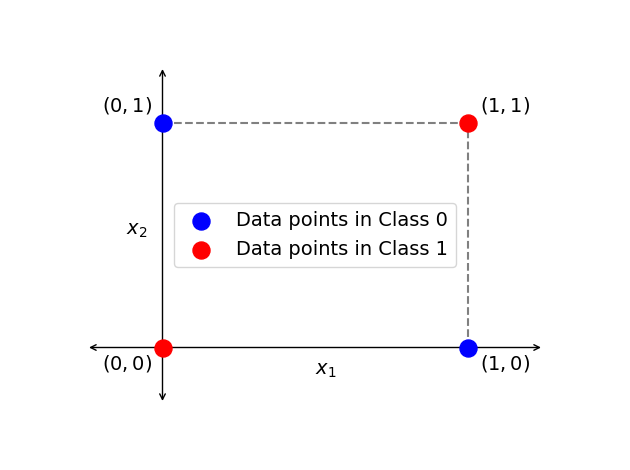}
    }
    \subfigure[Case 9]{
        \includegraphics[scale=0.36,trim={70px 55px 70px 55px}, clip]{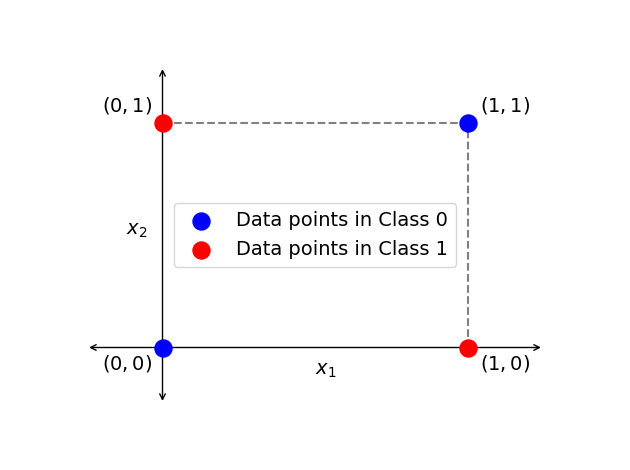}
    }
    \subfigure[Case 10]{
        \includegraphics[scale=0.36,trim={70px 55px 70px 55px}, clip]{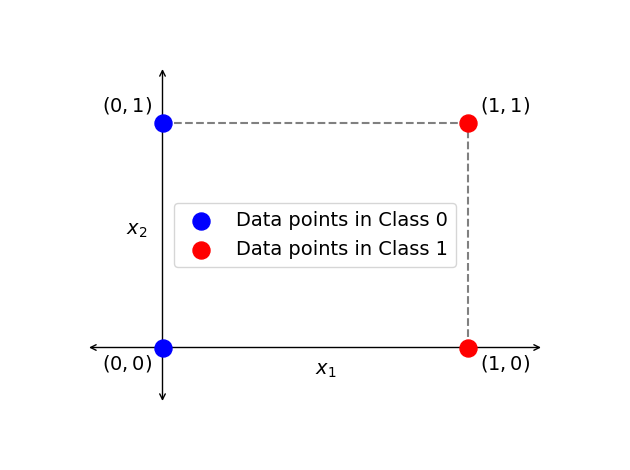}
    }
    \subfigure[Case 11]{
        \includegraphics[scale=0.36,trim={70px 55px 70px 55px}, clip]{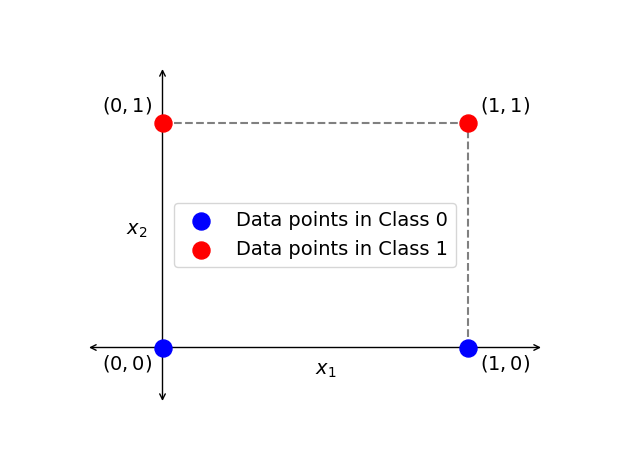}
    }
    \subfigure[Case 12]{
        \includegraphics[scale=0.36,trim={70px 55px 70px 55px}, clip]{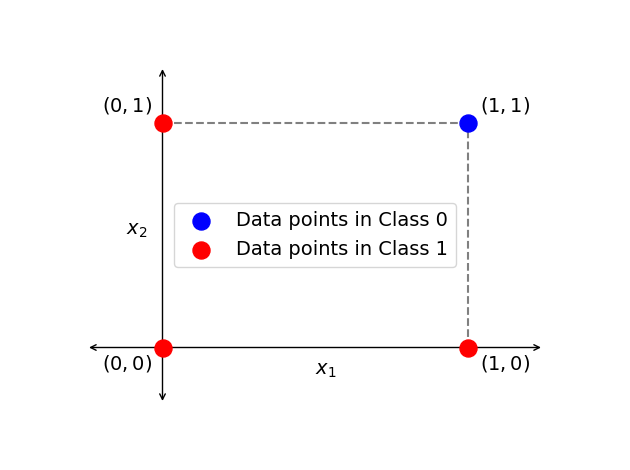}
    }
    \subfigure[Case 13]{
        \includegraphics[scale=0.35,trim={70px 55px 70px 55px}, clip]{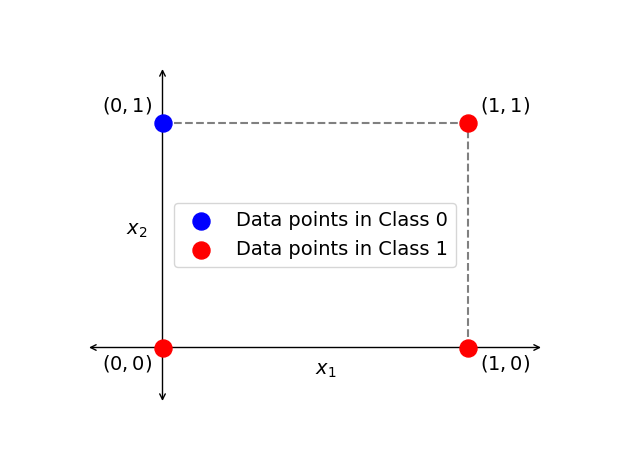}
    }
    \subfigure[Case 14]{
        \includegraphics[scale=0.35,trim={70px 55px 70px 55px}, clip]{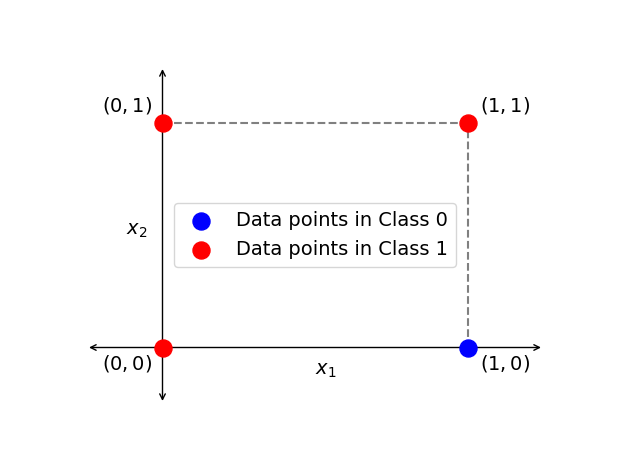}
    }
    \subfigure[Case 15]{
        \includegraphics[scale=0.35,trim={70px 55px 70px 55px}, clip]{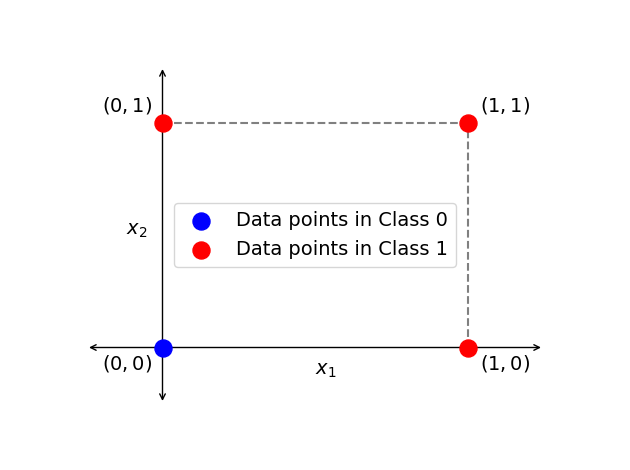}
    }
    \subfigure[Case 16]{
        \includegraphics[scale=0.35,trim={70px 55px 70px 55px}, clip]{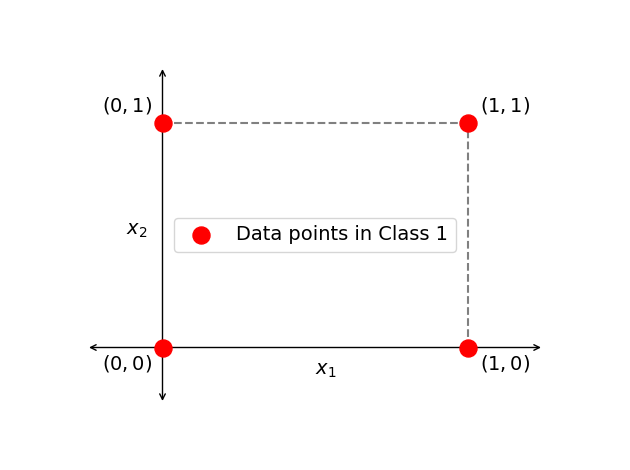}
    }
    \caption{16 cases of the 2-bit binary classification problem.}
    \label{fig:2-bit-classification-cases}
\end{figure}

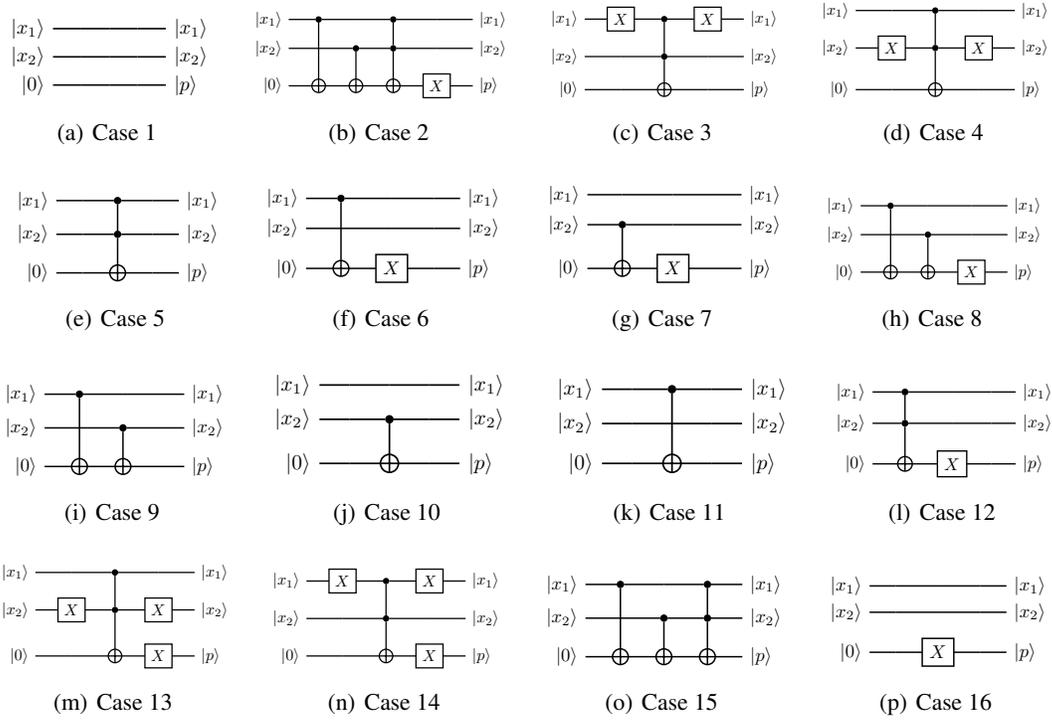
\begin{figure}[t!]
    \centering
    \subfigure[Case 1]{
        \begin{tikzpicture}
            \node[scale=0.75] {
            \begin{quantikz}
                \lstick{\ket{x_1}}  & \qw & \qw & \qw & \rstick{\ket{x_1}} \qw \\
                \lstick{\ket{x_2}}  & \qw & \qw & \qw & \rstick{\ket{x_2}} \qw \\
                \lstick{\ket{0}}    & \qw & \qw & \qw & \rstick{\ket{p}} \qw
            \end{quantikz}
            };
        \end{tikzpicture}
    }
    \subfigure[Case 2]{
        \begin{tikzpicture}
            \node[scale=0.6] {
            \begin{quantikz}
                \lstick{\ket{x_1}}  & \ctrl{2}  & \qw       & \ctrl{1}  & \qw       & \rstick{\ket{x_1}} \qw \\
                \lstick{\ket{x_2}}  & \qw       & \ctrl{1}  & \ctrl{1}  & \qw       & \rstick{\ket{x_2}} \qw \\
                \lstick{\ket{0}}    & \targ{}   & \targ{}   & \targ{}   & \gate{X}  & \rstick{\ket{p}} \qw
            \end{quantikz}
            };
        \end{tikzpicture}
    }
    \subfigure[Case 3]{
        \begin{tikzpicture}
            \node[scale=0.6] {
            \begin{quantikz}
                \lstick{\ket{x_1}}  & \gate{X}  & \ctrl{1}  & \gate{X}   & \rstick{\ket{x_1}} \qw \\
                \lstick{\ket{x_2}}  & \qw       & \ctrl{1}  & \qw        & \rstick{\ket{x_2}} \qw \\
                \lstick{\ket{0}}    & \qw       & \targ{}   & \qw        & \rstick{\ket{p}} \qw
            \end{quantikz}
            };
        \end{tikzpicture}
    }
    \subfigure[Case 4]{
        \begin{tikzpicture}
            \node[scale=0.6] {
            \begin{quantikz}
                \lstick{\ket{x_1}}  & \qw       & \ctrl{1} & \qw        & \rstick{\ket{x_1}} \qw \\
                \lstick{\ket{x_2}}  & \gate{X}  & \ctrl{1} & \gate{X}   & \rstick{\ket{x_2}} \qw \\
                \lstick{\ket{0}}    & \qw       & \targ{} & \qw        & \rstick{\ket{p}} \qw
            \end{quantikz}
            };
        \end{tikzpicture}
    }
    \subfigure[Case 5]{
        \begin{tikzpicture}
            \node[scale=0.7] {
            \begin{quantikz}
                \lstick{\ket{x_1}}  & \qw & \ctrl{1}  & \qw & \rstick{\ket{x_1}} \qw \\
                \lstick{\ket{x_2}}  & \qw & \ctrl{1}  & \qw & \rstick{\ket{x_2}} \qw \\
                \lstick{\ket{0}}    & \qw & \targ{}   & \qw & \rstick{\ket{p}} \qw
            \end{quantikz}
            };
        \end{tikzpicture} 
    }
    \subfigure[Case 6]{
        \begin{tikzpicture}
            \node[scale=0.7] {
            \begin{quantikz}
                \lstick{\ket{x_1}}  & \ctrl{2}  & \qw       & \qw & \rstick{\ket{x_1}} \qw \\
                \lstick{\ket{x_2}}  & \qw       & \qw       & \qw & \rstick{\ket{x_2}} \qw \\
                \lstick{\ket{0}}    & \targ{}   & \gate{X}  & \qw & \rstick{\ket{p}} \qw
            \end{quantikz}
            };
        \end{tikzpicture}
    }
    \subfigure[Case 7]{
        \begin{tikzpicture}
            \node[scale=0.7] {
            \begin{quantikz}
                \lstick{\ket{x_1}}  & \qw       & \qw       & \qw & \rstick{\ket{x_1}} \qw \\
                \lstick{\ket{x_2}}  & \ctrl{1}  & \qw       & \qw & \rstick{\ket{x_2}} \qw \\
                \lstick{\ket{0}}    & \targ{}   & \gate{X}  & \qw & \rstick{\ket{p}} \qw
            \end{quantikz}
            };
        \end{tikzpicture}
    }
    \subfigure[Case 8]{
        \begin{tikzpicture}
            \node[scale=0.6] {
            \begin{quantikz}
                \lstick{\ket{x_1}}  & \ctrl{2}  & \qw       & \qw       & \rstick{\ket{x_1}} \qw \\
                \lstick{\ket{x_2}}  & \qw       & \ctrl{1}  & \qw       & \rstick{\ket{x_2}} \qw \\
                \lstick{\ket{0}}    & \targ{}   & \targ{}   & \gate{X}  & \rstick{\ket{p}} \qw
            \end{quantikz}
            };
        \end{tikzpicture}
    }
    \subfigure[Case 9]{
        \begin{tikzpicture}
            \node[scale=0.7] {
            \begin{quantikz}
                \lstick{\ket{x_1}}  & \ctrl{2}  & \qw       & \qw & \rstick{\ket{x_1}} \qw \\
                \lstick{\ket{x_2}}  & \qw       & \ctrl{1}  & \qw & \rstick{\ket{x_2}} \qw \\
                \lstick{\ket{0}}    & \targ{}   & \targ{}   & \qw & \rstick{\ket{p}} \qw
            \end{quantikz}
            };
        \end{tikzpicture}
    }
    \subfigure[Case 10]{
        \begin{tikzpicture}
            \node[scale=0.8] {
            \begin{quantikz}
                \lstick{\ket{x_1}}  & \qw & \qw       & \qw & \rstick{\ket{x_1}} \qw \\
                \lstick{\ket{x_2}}  & \qw & \ctrl{1}  & \qw & \rstick{\ket{x_2}} \qw \\
                \lstick{\ket{0}}    & \qw & \targ{}   & \qw & \rstick{\ket{p}} \qw
            \end{quantikz}
            };
        \end{tikzpicture}
    }
    \subfigure[Case 11]{
        \begin{tikzpicture}
            \node[scale=0.8] {
            \begin{quantikz}
                \lstick{\ket{x_1}}  & \qw & \ctrl{2}  & \qw & \rstick{\ket{x_1}} \qw \\
                \lstick{\ket{x_2}}  & \qw & \qw       & \qw & \rstick{\ket{x_2}} \qw \\
                \lstick{\ket{0}}    & \qw & \targ{}   & \qw & \rstick{\ket{p}} \qw
            \end{quantikz}
            };
        \end{tikzpicture}
    }
    \subfigure[Case 12]{
        \begin{tikzpicture}
            \node[scale=0.65] {
            \begin{quantikz}
                \lstick{\ket{x_1}}  & \ctrl{1}  & \qw       & \qw & \rstick{\ket{x_1}} \qw \\
                \lstick{\ket{x_2}}  & \ctrl{1}  & \qw       & \qw & \rstick{\ket{x_2}} \qw \\
                \lstick{\ket{0}}    & \targ{}   & \gate{X}  & \qw & \rstick{\ket{p}} \qw
            \end{quantikz}
            };
        \end{tikzpicture}
    }
    \subfigure[Case 13]{
        \begin{tikzpicture}
            \node[scale=0.6] {
            \begin{quantikz}
                \lstick{\ket{x_1}}  & \qw       & \ctrl{1}  & \qw       & \rstick{\ket{x_1}} \qw \\
                \lstick{\ket{x_2}}  & \gate{X}  & \ctrl{1}  & \gate{X}  & \rstick{\ket{x_2}} \qw \\
                \lstick{\ket{0}}    & \qw       & \targ{}   & \gate{X}  & \rstick{\ket{p}} \qw
            \end{quantikz}
            };
        \end{tikzpicture}
    }
    \subfigure[Case 14]{
        \begin{tikzpicture}
            \node[scale=0.6] {
            \begin{quantikz}
                \lstick{\ket{x_1}}  & \gate{X}  & \ctrl{1}  & \gate{X}   & \rstick{\ket{x_1}} \qw \\
                \lstick{\ket{x_2}}  & \qw       & \ctrl{1}  & \qw        & \rstick{\ket{x_2}} \qw \\
                \lstick{\ket{0}}    & \qw       & \targ{}   & \gate{X}   & \rstick{\ket{p}} \qw
            \end{quantikz}
            };
        \end{tikzpicture}
    }
    \subfigure[Case 15]{
        \begin{tikzpicture}
            \node[scale=0.7] {
            \begin{quantikz}
                \lstick{\ket{x_1}}  & \ctrl{2}  & \qw       & \ctrl{1}  & \rstick{\ket{x_1}} \qw \\
                \lstick{\ket{x_2}}  & \qw       & \ctrl{1}  & \ctrl{1}  & \rstick{\ket{x_2}} \qw \\
                \lstick{\ket{0}}    & \targ{}   & \targ{}   & \targ{}   & \rstick{\ket{p}} \qw
            \end{quantikz}
            };
        \end{tikzpicture}
    }
    \subfigure[Case 16]{
        \begin{tikzpicture}
            \node[scale=0.7] {
            \begin{quantikz}
                \lstick{\ket{x_1}}  & \qw & \qw       & \qw & \rstick{\ket{x_1}} \qw \\
                \lstick{\ket{x_2}}  & \qw & \qw       & \qw & \rstick{\ket{x_2}} \qw \\
                \lstick{\ket{0}}    & \qw & \gate{X}  & \qw & \rstick{\ket{p}} \qw
            \end{quantikz}
            };
        \end{tikzpicture}
    }
    \caption{Quantum circuits for each case of the 2-bit binary classification problem.}
    \label{fig:2-bit-classification-circuits}
\end{figure}

To demonstrate a proof of concept for the proposed quantum discriminator, we use it to classify 2-dimensional binary data points.
There are four such data points: $(0,0)$, $(0,1)$, $(1,0)$ and $(1,1)$, located on the corners of a unit square.
There are $2^4 = 16$ ways of classifying these points into two classes as shown in Figure \ref{fig:2-bit-classification-cases}.
For example, one such way could be: $(0,0)$ and $(0,1)$ belong to Class $0$, and $(1,0)$ and $(1,1)$ belong to Class $1$, as shown by Case 10 in Figure \ref{fig:2-bit-classification-cases}.
In Figure \ref{fig:2-bit-classification-circuits}, we show the quantum circuits that can classify each of the $16$ cases in Figure \ref{fig:2-bit-classification-cases}.
In Table \ref{tab:2-bit-classification-unitaries}, we present the unitary matrices of the quantum discriminator models that classify each of the $16$ cases in Figure \ref{fig:2-bit-classification-cases}.

We now elaborate Case $4$ from Figure \ref{fig:2-bit-classification-cases}.
In Case $4$, the points $(0,0)$, $(0,1)$ and $(1,1)$ belong to Class $0$, while the point $(1,0)$ belongs to Class $1$.
The corresponding quantum circuit (Case 4 in Figure \ref{fig:2-bit-classification-circuits}) used for classification takes as input the two quantum data features ($\ket{x_1}$ and $\ket{x_2}$) as well as the prediction qubit initialized to $\ket{0}$.
Next, it applies the Pauli-$X$ gate  on $\ket{x_2}$, followed by a Toffoli gate (also called the CCNOT gate) with $\ket{x_1}$ and $\ket{x_2}$ as the control bits and $\ket{0}$ as the target bit, followed by another Pauli-$X$ gate on $\ket{x_2}$.

The corresponding unitary operator ($U_\Theta$) is shown in the Case 4 of Table \ref{tab:2-bit-classification-unitaries}. 
It looks like an identity matrix with one caveat.
The $2 \times 2$ sub-matrix starting at the fifth row and fifth column is a Pauli-$X$ operator instead of a $2 \times 2$ identity matrix.
Note that the first and second rows/columns of all $U_\Theta$ shown in Tables \ref{tab:2-bit-classification-unitaries} correspond to the data point $(0,0)$.
Similarly, third and fourth rows/columns correspond to the data point $(0,1)$, fifth and sixth rows/columns correspond to $(1,0)$, and seventh and eighth rows/columns correspond to $(1,1)$.
The starting indices ($1$, $3$, $5$ and $7$) of these $2 \times 2$ sub-matrices can be computed from the equation on line $7$ of Algorithm \ref{algo:training}.
For each data point in each of the $16$ cases, we can independently update the classification operator ($U_\Theta$) so that it correctly classifies the said data point.
In this way, the quantum discriminator is able to achieve near perfect accuracies on binary classification problems, provided appropriate binary features have been extracted from the data.

\begin{table}[t!]
    \centering
    \caption{Unitary matrices for each case of the 2-bit binary classification problem.}
    \label{tab:2-bit-classification-unitaries}
    \vspace{10pt}
    \begin{tabular}{c c c c}
        \tiny{$\begin{bmatrix}
             1 &  0 &  0 &  0 &  0 &  0 &  0 &  0 \\
             0 &  1 &  0 &  0 &  0 &  0 &  0 &  0 \\
             0 &  0 &  1 &  0 &  0 &  0 &  0 &  0 \\
             0 &  0 &  0 &  1 &  0 &  0 &  0 &  0 \\
             0 &  0 &  0 &  0 &  1 &  0 &  0 &  0 \\
             0 &  0 &  0 &  0 &  0 &  1 &  0 &  0 \\
             0 &  0 &  0 &  0 &  0 &  0 &  1 &  0 \\
             0 &  0 &  0 &  0 &  0 &  0 &  0 &  1 \\
        \end{bmatrix}$}
        & 
        \tiny{$\begin{bmatrix}
             0 &  1 &  0 &  0 &  0 &  0 &  0 &  0 \\
             1 &  0 &  0 &  0 &  0 &  0 &  0 &  0 \\
             0 &  0 &  1 &  0 &  0 &  0 &  0 &  0 \\
             0 &  0 &  0 &  1 &  0 &  0 &  0 &  0 \\
             0 &  0 &  0 &  0 &  1 &  0 &  0 &  0 \\
             0 &  0 &  0 &  0 &  0 &  1 &  0 &  0 \\
             0 &  0 &  0 &  0 &  0 &  0 &  1 &  0 \\
             0 &  0 &  0 &  0 &  0 &  0 &  0 &  1 \\
        \end{bmatrix}$}
        &
        \tiny{$\begin{bmatrix}
             1 &  0 &  0 &  0 &  0 &  0 &  0 &  0 \\
             0 &  1 &  0 &  0 &  0 &  0 &  0 &  0 \\
             0 &  0 &  0 &  1 &  0 &  0 &  0 &  0 \\
             0 &  0 &  1 &  0 &  0 &  0 &  0 &  0 \\
             0 &  0 &  0 &  0 &  1 &  0 &  0 &  0 \\
             0 &  0 &  0 &  0 &  0 &  1 &  0 &  0 \\
             0 &  0 &  0 &  0 &  0 &  0 &  1 &  0 \\
             0 &  0 &  0 &  0 &  0 &  0 &  0 &  1 \\
        \end{bmatrix}$}
        & 
        \tiny{$\begin{bmatrix}
            1 &  0 &  0 &  0 &  0 &  0 &  0 &  0 \\
             0 &  1 &  0 &  0 &  0 &  0 &  0 &  0 \\
             0 &  0 &  1 &  0 &  0 &  0 &  0 &  0 \\
             0 &  0 &  0 &  1 &  0 &  0 &  0 &  0 \\
             0 &  0 &  0 &  0 &  0 &  1 &  0 &  0 \\
             0 &  0 &  0 &  0 &  1 &  0 &  0 &  0 \\
             0 &  0 &  0 &  0 &  0 &  0 &  1 &  0 \\
             0 &  0 &  0 &  0 &  0 &  0 &  0 &  1 \\
        \end{bmatrix}$} 
        \\
        Case 1 & Case 2 & Case 3 & Case 4 \\
        \noalign{\smallskip}\\
        \tiny{$\begin{bmatrix}
             1 &  0 &  0 &  0 &  0 &  0 &  0 &  0 \\
             0 &  1 &  0 &  0 &  0 &  0 &  0 &  0 \\
             0 &  0 &  1 &  0 &  0 &  0 &  0 &  0 \\
             0 &  0 &  0 &  1 &  0 &  0 &  0 &  0 \\
             0 &  0 &  0 &  0 &  1 &  0 &  0 &  0 \\
             0 &  0 &  0 &  0 &  0 &  1 &  0 &  0 \\
             0 &  0 &  0 &  0 &  0 &  0 &  0 &  1 \\
             0 &  0 &  0 &  0 &  0 &  0 &  1 &  0 \\
        \end{bmatrix}
        $}
        &
        \tiny{$
        \begin{bmatrix}
             0 &  1 &  0 &  0 &  0 &  0 &  0 &  0 \\
             1 &  0 &  0 &  0 &  0 &  0 &  0 &  0 \\
             0 &  0 &  0 &  1 &  0 &  0 &  0 &  0 \\
             0 &  0 &  1 &  0 &  0 &  0 &  0 &  0 \\
             0 &  0 &  0 &  0 &  1 &  0 &  0 &  0 \\
             0 &  0 &  0 &  0 &  0 &  1 &  0 &  0 \\
             0 &  0 &  0 &  0 &  0 &  0 &  1 &  0 \\
             0 &  0 &  0 &  0 &  0 &  0 &  0 &  1 \\
        \end{bmatrix}
        $}
        &
        \tiny{$
        \begin{bmatrix}
             0 &  1 &  0 &  0 &  0 &  0 &  0 &  0 \\
             1 &  0 &  0 &  0 &  0 &  0 &  0 &  0 \\
             0 &  0 &  1 &  0 &  0 &  0 &  0 &  0 \\
             0 &  0 &  0 &  1 &  0 &  0 &  0 &  0 \\
             0 &  0 &  0 &  0 &  0 &  1 &  0 &  0 \\
             0 &  0 &  0 &  0 &  1 &  0 &  0 &  0 \\
             0 &  0 &  0 &  0 &  0 &  0 &  1 &  0 \\
             0 &  0 &  0 &  0 &  0 &  0 &  0 &  1 \\
        \end{bmatrix}
        $}
        &
        \tiny{$\begin{bmatrix}
             0 &  1 &  0 &  0 &  0 &  0 &  0 &  0 \\
             1 &  0 &  0 &  0 &  0 &  0 &  0 &  0 \\
             0 &  0 &  1 &  0 &  0 &  0 &  0 &  0 \\
             0 &  0 &  0 &  1 &  0 &  0 &  0 &  0 \\
             0 &  0 &  0 &  0 &  1 &  0 &  0 &  0 \\
             0 &  0 &  0 &  0 &  0 &  1 &  0 &  0 \\
             0 &  0 &  0 &  0 &  0 &  0 &  0 &  1 \\
             0 &  0 &  0 &  0 &  0 &  0 &  1 &  0 \\
        \end{bmatrix}
        $}
        \\
        Case 5 & Case 6 & Case 7 & Case 8 \\
        \noalign{\smallskip}\\
        \tiny{$\begin{bmatrix}
             1 &  0 &  0 &  0 &  0 &  0 &  0 &  0 \\
             0 &  1 &  0 &  0 &  0 &  0 &  0 &  0 \\
             0 &  0 &  0 &  1 &  0 &  0 &  0 &  0 \\
             0 &  0 &  1 &  0 &  0 &  0 &  0 &  0 \\
             0 &  0 &  0 &  0 &  0 &  1 &  0 &  0 \\
             0 &  0 &  0 &  0 &  1 &  0 &  0 &  0 \\
             0 &  0 &  0 &  0 &  0 &  0 &  1 &  0 \\
             0 &  0 &  0 &  0 &  0 &  0 &  0 &  1 \\
        \end{bmatrix}
        $}
        &
        \tiny{$\begin{bmatrix}
             1 &  0 &  0 &  0 &  0 &  0 &  0 &  0 \\
             0 &  1 &  0 &  0 &  0 &  0 &  0 &  0 \\
             0 &  0 &  0 &  1 &  0 &  0 &  0 &  0 \\
             0 &  0 &  1 &  0 &  0 &  0 &  0 &  0 \\
             0 &  0 &  0 &  0 &  1 &  0 &  0 &  0 \\
             0 &  0 &  0 &  0 &  0 &  1 &  0 &  0 \\
             0 &  0 &  0 &  0 &  0 &  0 &  0 &  1 \\
             0 &  0 &  0 &  0 &  0 &  0 &  1 &  0 \\
        \end{bmatrix}
        $}
        &
        \tiny{$
        \begin{bmatrix}
             1 &  0 &  0 &  0 &  0 &  0 &  0 &  0 \\
             0 &  1 &  0 &  0 &  0 &  0 &  0 &  0 \\
             0 &  0 &  1 &  0 &  0 &  0 &  0 &  0 \\
             0 &  0 &  0 &  1 &  0 &  0 &  0 &  0 \\
             0 &  0 &  0 &  0 &  0 &  1 &  0 &  0 \\
             0 &  0 &  0 &  0 &  1 &  0 &  0 &  0 \\
             0 &  0 &  0 &  0 &  0 &  0 &  0 &  1 \\
             0 &  0 &  0 &  0 &  0 &  0 &  1 &  0 \\
        \end{bmatrix}
        $}
        &
        \tiny{$\begin{bmatrix}
             0 &  1 &  0 &  0 &  0 &  0 &  0 &  0 \\
             1 &  0 &  0 &  0 &  0 &  0 &  0 &  0 \\
             0 &  0 &  0 &  1 &  0 &  0 &  0 &  0 \\
             0 &  0 &  1 &  0 &  0 &  0 &  0 &  0 \\
             0 &  0 &  0 &  0 &  0 &  1 &  0 &  0 \\
             0 &  0 &  0 &  0 &  1 &  0 &  0 &  0 \\
             0 &  0 &  0 &  0 &  0 &  0 &  1 &  0 \\
             0 &  0 &  0 &  0 &  0 &  0 &  0 &  1 \\
        \end{bmatrix}
        $}
        \\
        Case 9 & Case 10 & Case 11 & Case 12 \\
        \noalign{\smallskip}\\
        \tiny{$\begin{bmatrix}
             0 &  1 &  0 &  0 &  0 &  0 &  0 &  0 \\
             1 &  0 &  0 &  0 &  0 &  0 &  0 &  0 \\
             0 &  0 &  0 &  1 &  0 &  0 &  0 &  0 \\
             0 &  0 &  1 &  0 &  0 &  0 &  0 &  0 \\
             0 &  0 &  0 &  0 &  1 &  0 &  0 &  0 \\
             0 &  0 &  0 &  0 &  0 &  1 &  0 &  0 \\
             0 &  0 &  0 &  0 &  0 &  0 &  0 &  1 \\
             0 &  0 &  0 &  0 &  0 &  0 &  1 &  0 \\
        \end{bmatrix}
        $}
        &
        \tiny{$\begin{bmatrix}
             0 &  1 &  0 &  0 &  0 &  0 &  0 &  0 \\
             1 &  0 &  0 &  0 &  0 &  0 &  0 &  0 \\
             0 &  0 &  1 &  0 &  0 &  0 &  0 &  0 \\
             0 &  0 &  0 &  1 &  0 &  0 &  0 &  0 \\
             0 &  0 &  0 &  0 &  0 &  1 &  0 &  0 \\
             0 &  0 &  0 &  0 &  1 &  0 &  0 &  0 \\
             0 &  0 &  0 &  0 &  0 &  0 &  0 &  1 \\
             0 &  0 &  0 &  0 &  0 &  0 &  1 &  0 \\
        \end{bmatrix}
        $}
        &
        \tiny{$\begin{bmatrix}
             1 &  0 &  0 &  0 &  0 &  0 &  0 &  0 \\
             0 &  1 &  0 &  0 &  0 &  0 &  0 &  0 \\
             0 &  0 &  0 &  1 &  0 &  0 &  0 &  0 \\
             0 &  0 &  1 &  0 &  0 &  0 &  0 &  0 \\
             0 &  0 &  0 &  0 &  0 &  1 &  0 &  0 \\
             0 &  0 &  0 &  0 &  1 &  0 &  0 &  0 \\
             0 &  0 &  0 &  0 &  0 &  0 &  0 &  1 \\
             0 &  0 &  0 &  0 &  0 &  0 &  1 &  0 \\
        \end{bmatrix}
        $}
        &
        \tiny{$\begin{bmatrix}
             0 &  1 &  0 &  0 &  0 &  0 &  0 &  0 \\
             1 &  0 &  0 &  0 &  0 &  0 &  0 &  0 \\
             0 &  0 &  0 &  1 &  0 &  0 &  0 &  0 \\
             0 &  0 &  1 &  0 &  0 &  0 &  0 &  0 \\
             0 &  0 &  0 &  0 &  0 &  1 &  0 &  0 \\
             0 &  0 &  0 &  0 &  1 &  0 &  0 &  0 \\
             0 &  0 &  0 &  0 &  0 &  0 &  0 &  1 \\
             0 &  0 &  0 &  0 &  0 &  0 &  1 &  0 \\
        \end{bmatrix}
        $}
        \\
        Case 13 & Case 14 & Case 15 & Case 16 
    \end{tabular}
\end{table}

\end{document}